\documentclass[a4paper,11pt]{article}
\usepackage{jheppub} 
\usepackage{lineno}

\usepackage{combelow}
\usepackage{amsmath}
\usepackage{amssymb}

\makeatletter
\DeclareFontFamily{OMX}{MnSymbolE}{}
\DeclareSymbolFont{MnLargeSymbols}{OMX}{MnSymbolE}{m}{n}
\SetSymbolFont{MnLargeSymbols}{bold}{OMX}{MnSymbolE}{b}{n}
\DeclareFontShape{OMX}{MnSymbolE}{m}{n}{
    <-6>  MnSymbolE5
   <6-7>  MnSymbolE6
   <7-8>  MnSymbolE7
   <8-9>  MnSymbolE8
   <9-10> MnSymbolE9
  <10-12> MnSymbolE10
  <12->   MnSymbolE12
}{}
\DeclareFontShape{OMX}{MnSymbolE}{b}{n}{
    <-6>  MnSymbolE-Bold5
   <6-7>  MnSymbolE-Bold6
   <7-8>  MnSymbolE-Bold7
   <8-9>  MnSymbolE-Bold8
   <9-10> MnSymbolE-Bold9
  <10-12> MnSymbolE-Bold10
  <12->   MnSymbolE-Bold12
}{}

\let\llangle\@undefined
\let\rrangle\@undefined
\DeclareMathDelimiter{\llangle}{\mathopen}%
                     {MnLargeSymbols}{'164}{MnLargeSymbols}{'164}
\DeclareMathDelimiter{\rrangle}{\mathclose}%
                     {MnLargeSymbols}{'171}{MnLargeSymbols}{'171}
\makeatother

\newcommand{\avr}[1]{\left\langle #1 \right\rangle}
\newcommand{\bs}{\boldsymbol}
\DeclareMathOperator{\Tr}{Tr}

\usepackage{cancel}
\usepackage[normalem]{ulem}
\usepackage[dvipsnames]{xcolor}
\newcommand{\replaceR}[2]{{{\color{BrickRed}{#1}}{\color{NavyBlue}{\ifmmode\text{\sout{\ensuremath{#2}}}\else\sout{#2}\fi}}}}

\newcommand{\replaceB}[2]{{{\color{RedViolet}{#1}}{\color{BlueViolet}{\ifmmode\text{\sout{\ensuremath{#2}}}\else\sout{#2}\fi}}}}
\newcommand{\vb}[2]{{{\color{RedViolet}{#1}}{\color{BlueViolet}{\ifmmode\text{\sout{\ensuremath{#2}}}\else\sout{#2}\fi}}}}

\newcommand{\replaceC}[2]{{{\color{RedOrange}{#1}}{\color{OliveGreen}{\ifmmode\text{\sout{\ensuremath{#2}}}\else\sout{#2}\fi}}}}

\newcommand{\changeq}[3]{{\color{red} \ifmmode\text{\sout{\ensuremath{#1}}}\else\sout{#1}\fi}{\color[rgb]{0.56,0.0,1.0} #2}{\color{blue}[#3]}}

\newcommand{\aavr}[1]{\left \llangle #1 \right \rrangle}

\arxivnumber{2411.15085} 

\title{On the origin of mixed inhomogeneous phase in vortical gluon plasma}

\author[a]{V.~V.~Braguta,} 
\author[b,c]{M.~N.~Chernodub,}
\author[d]{Ya.~A.~Gershtein,}
\author[a]{A.~A.~Roenko}
\affiliation[a]{Bogoliubov Laboratory of Theoretical Physics, Joint Institute for Nuclear Research, 
Dubna, 141980 Russia}
\affiliation[b]{Institut Denis Poisson UMR 7013, Universit\'e de Tours, 
Tour, 37200 France}
\affiliation[c]{Department of Physics, West University of Timi\cb{s}oara, 
Bd.~Vasile P\^arvan 4, Timi\cb{s}oara, 300223 Romania}
\affiliation[d]{Moscow Institute of Physics and Technology, 
Dolgoprudny, 141700 Russia}

\emailAdd{roenko@theor.jinr.ru}

\abstract{
Recently, lattice simulations of SU(3) Yang-Mills theory revealed that rotating hot gluon matter in thermal equilibrium possesses a novel inhomogeneous phase consisting of the deconfinement phase located in the center region, which is spatially separated from the confinement phase in the periphery. This inhomogeneous two-phase structure is also expected to be produced by vorticity in quark-gluon plasma formed in non-central relativistic heavy-ion collisions. We show that its vortical properties are determined by two types of couplings of the angular velocity to the gluon fields: a linear coupling to the mechanical angular momentum of gluons and a quadratic ``magnetovortical'' coupling to a chromomagnetic component. We demonstrate numerically that the distinctive inhomogeneous structure of the vortical (quark-)gluon plasma is determined by the latter, while the former plays only a subleading role. We argue that the anisotropy of the gluonic action in the curved co-rotating background can quantitatively explain the remarkable property that the spatial structure of this inhomogeneous phase disobeys the picture based on a straightforward implementation of the Tolman-Ehrenfest law. We also support our findings with Monte Carlo simulations of Yang-Mills plasma at the real-valued angular frequency, which take into account only the magnetic part of the action.
}

\begin{document}
\maketitle
\flushbottom

\section{Introduction}\label{sec_Intro}

The influence of rapid rotation on the properties of different physical systems proved to be an intriguing and important area of scientific research. There are many examples when rotation plays a significant role in astrophysics~\cite{Watts:2016uzu, Grenier:2015pya}, condensed matter physics~\cite{Basar:2013iaa, Landsteiner:2013sja}, nuclear physics~\cite{PhysRevLett.122.177202}, 
to mention a few. The fastest rotation in Nature is realized in heavy-ion collision experiments that create highly vortical quark-gluon matter~\cite{Jiang:2016woz, Baznat:2013zx}. The global polarization of $\Lambda$ and $\bar \Lambda$ hyperons obtained by the STAR Collaboration indicate that the vorticity of hot quark-gluon matter in noncentral collisions reaches the enormous value $\sim 10^{22}$~Hz ($\Omega \sim 10$~MeV)~\cite{STAR:2017ckg}. In terms of speed of light $c$, the mechanical vortical motion of plasma acquires significant velocities $v \sim \Omega r \sim (0.1 \dots 0.3) c$ at typical distances $r \sim \mbox{(a few)}\times \mbox{fm}$ within the expanding fireball. 

It is reasonable to assume that the high vorticity of the quark-gluon matter affects its thermodynamic properties and, in particular, modifies the phase diagram of QCD. The last question has been addressed in various papers within different theoretical approaches (see, for instance,~\cite{Chernodub:2016kxh, Jiang:2016wvv, Chernodub:2017ref, Wang:2018sur, Chen:2020ath,  Golubtsova:2022ldm, Chen:2022smf, Singha:2024tpo, Jiang:2023zzu, Sun:2023kuu, Chen:2023cjt, Zhao:2022uxc, Yadav:2022qcl, Braga:2022yfe, Mehr:2022tfq, Sadooghi:2021upd, Fujimoto:2021xix, Zhang:2020hha, Chen:2022mhf}). Most of the theoretical works predict a decrease in the critical temperatures of deconfinement and chiral transitions in rotating hot QCD matter. On the contrary, some theoretical works point out to quite the opposite behaviour~\cite{Jiang:2021izj, Mameda:2023sst, Chen:2024jet, Sun:2024anu}. The influence of rotation on the phase transitions in gluodynamics and QCD within lattice simulation has also been studied, following an early work of Ref.~\cite{Yamamoto:2013zwa}, in several papers both in pure gluonic plasmas~\cite{Braguta:2020biu, Braguta:2021jgn, Braguta:2021ucr, Chernodub:2022veq} as well as in lattice QCD with dynamical quarks~\cite{Braguta:2022str, Yang:2023vsw}. Contrary to the predictions of many theoretical approaches, these first-principles simulations demonstrate that both transition temperatures rise with increasing rotation.

It was also suggested that rotation could lead to a new, inhomogeneous phase in QCD, in which spatially separated hadronic and plasma phases could exist simultaneously~\cite{Chernodub:2020qah}. This result, obtained in a low-dimensional confining model, has been subsequently supported by a study in the holography approach~\cite{Braga:2023qej}. Later, the first-principle lattice simulation conducted in Ref.~\cite{Braguta:2023iyx} confirmed that mixed-phase indeed takes place in $\textrm{SU}(3)$ gluodynamics. It was shown that the local critical temperature increases and the mixed state has an inverted phase arrangement compared to Refs.~\cite{Chernodub:2020qah, Braga:2023qej}. In particular, the lattice simulations of Ref.~\cite{Braguta:2023iyx} have revealed that for the temperatures close to the critical one, there appears a mixed state with deconfinement at the center and confinement close to the boundaries of the studied volume. This observation has also been supported in analytical approaches~\cite{Chen:2024tkr, Jiang:2024zsw}. 

In a parallel development, the rotating gluon plasma has numerically been shown to possess a negative moment of inertia in a region of temperatures above the deconfining phase transition $T_c$ up to the ``supervortical temperature'' $T_s \simeq 1.5 T_c$~\cite{Braguta:2023yjn, Braguta:2023kwl, Braguta:2023qex}. This phenomenon has been proposed to originate from the negative Barnett effect and the thermal evaporation of the non-perturbative chromomagnetic condensate, which gives, together with the standard angular momentum, a contribution to the mechanical moment of inertia of the gluon plasma~\cite{Braguta:2023tqz}. Similar features were also argued to emerge in a relativistic bosonic gas under certain regimes~\cite{Siri:2024scq, Siri:2024cjw}.

In this paper, we present a more detailed study of the mixed phase found in Ref.~\cite{Braguta:2023iyx} in an inquiry to shed more light on its origin. In the spirit of Ref.~\cite{Braguta:2023tqz}, we separate the effects coming from the mechanical moment of thermal gluons from the chromomagnetic contribution and analyze their impact on the formation of the inhomogeneous plasma.

This paper is organized as follows. In Section~\ref{sec_Couplings}, we discuss the kinematical properties of rigid rotation at the level of the QCD Lagrangian. We stress the non-linear nature of the coupling of the angular velocity to the gluonic fields as opposed to the linear coupling of the rotation to quarks. A special role of the chromomagnetic degrees of freedom is highlighted. In Section~\ref{sec_Formulation}, we present a lattice framework implemented in the rotating $\textrm{SU}(3)$ gluodynamics and discuss the peculiarities of the rotating system. Section~\ref{sec_Coexistence} is devoted to the results of the lattice simulations revealing the spatial distribution of the Polyakov loop. This Section also provides details on the calculation of the local critical temperature of the deconfinement transition. In Section~\ref{sec_Decomposition}, we study how the terms in lattice action, which are linear/quadratic in angular velocity, influence the critical temperature and discuss the analytic continuation of our results from the domain of imaginary angular frequencies implemented in the lattice simulations, to the real-valued frequencies, relevant to the realistic plasmas. In Section~\ref{sec_Local}, we show that the metric of the rotating reference frame gives rise to asymmetry in the coupling constants in the gluon action. We demonstrate that it is the peculiar effect of metric that leads to the shift of the critical temperature in the periphery regions compared to the central regions. The last Section~\ref{sec_Conclusions} is devoted to the conclusions and detailed discussions of the numerical results. Relevant technical details are provided in the Appendix.

\section{Gluons in rotation: Mechanical and magnetic vortical couplings}
\label{sec_Couplings}

\subsection{Kinematics of rigid rotation}

The thermodynamic equilibrium of a rotating fluid is achieved in a state of a rigid rotation when the angular velocity of every fluid element has the same value $\boldsymbol{\Omega}$ about the axis of rotation ${\boldsymbol{n}} = {\boldsymbol{\Omega}}/\Omega$ with the angular frequency $\Omega = |\boldsymbol{\Omega}|$. The solid rotation excludes the hydrodynamic friction of the adjacent fluid layers and prevents the generation of heat (entropy) in consistency with the notion of thermodynamic equilibrium~\cite{Israel:1976tn, cercignani02}. 

It is convenient to describe the rotating system in a co-rotating reference frame, $x^\mu \equiv (x^0, x^1, x^2, x^3) = (t, x, y, z)$, in which the rigidly rotating system appears static. Due to the symmetry of the problem, we consider a system bounded by a cylinder of the radius $R$ and briefly use the cylindrical coordinates, $x^\mu = (t,  r  \sin \varphi,  r  \cos \varphi, z)$, with $\varphi$, $ r $, and $z$ being the azimuthal angle, the radial and longitudinal coordinates, respectively. The system rotates about the $z$ axis, ${\boldsymbol{\Omega}} = \Omega {\bf{e}}_z$.

The metric in the co-rotating system can be obtained by a coordinate transformation from the laboratory frame to the co-rotating frame. To this end,  we denote by the subscript ``lab'' the coordinates in the flat Minkowski metric of a laboratory frame, $x^\mu_{\rm lab}$. The only difference between the laboratory and co-rotating frames appears in the azimuthal coordinate $\varphi = [\varphi_{\rm lab} - \Omega t]_{2\pi}$, where $[\dots]_{2\pi}$ denote taking a modulo of $2\pi$. The other coordinate components coincide in both frames, $t = t_{\rm lab}$, $r = r_{\rm lab}$ and $z = z_{\rm lab}$. The diffeomorphism gives us the metric in the co-rotating frame in the Cartesian coordinates:
\begin{equation}\label{eq_metric}
g_{\mu \nu} = 
\begin{pmatrix}
1 -  r ^2 \Omega^2 & \Omega y & -\Omega x & 0 \\
\Omega y & -1 & 0 & 0  \\ 
-\Omega x & 0 & -1 & 0 \\
0 & 0 & 0 & -1
\end{pmatrix}\,,
\end{equation}
where $ r = \sqrt{x^2+y^2}$ is the distance from the rotation axis. The co-rotating coordinates correspond to a non-inertial nature contrary to the inertial laboratory frame. Notice that $g \equiv {\rm det}\, g_{\mu\nu} = - 1$.

In a relativistic theory, the solid rotation can only be considered for a physical body lying entirely within a light cylinder. It means that the farthermost point of the body from the rotation axis should possess a linear velocity smaller than the speed of light, $R \Omega < 1$. This requirement prevents the system from breaking the fundamental causality constraint. For the causality-respecting systems, the $g_{00}$ component of the co-rotating metric~\eqref{eq_metric} is a non-negative quantity at any point of the rotating body.

\subsection{QCD matter in rotating frame}

\subsubsection{Fermions}

Before going into the discussion of the main subject of this paper, the kinematics of rotating gluons, it is convenient to discuss first how rigid rotation affects the kinematics of fermions. The Lagrangian of a Dirac particle in the co-rotating frame corresponds to the Dirac Lagrangian in the curved spacetime described by the metric~\eqref{eq_metric}. Without repeating the details of the derivation that can be found elsewhere (for example, in Refs.~\cite{Ambrus:2015lfr, Jiang:2016wvv, Chernodub:2016kxh}), we give here the Dirac Lagrangian written the co-rotating coordinates
\begin{align}
{\cal L}_{\psi} = {\cal L}^{(0)}_{\psi} + {\cal L}^{(1)}_{\psi}
               \equiv {\cal L}_{\psi,{\rm lab}} + {\cal L}_{\psi,{\rm mech}}\,,
\label{eq_L_psi}
\end{align}            
which is a linear function of the angular frequency $\Omega$ with ${\cal L}^{(n)}_{\psi} \propto \Omega^n$ ($n=0,1$). The terms in Lagrangian~\eqref{eq_L_psi} are
\begin{align}
     {\cal L}_{\psi,{\rm lab}} & \equiv {\cal L}^{(0)}_{\psi} = {\bar \psi} \bigl(i \gamma^\mu \partial_\mu - m\bigr) \psi\,, 
     \label{eq_L_psi_0}\\
     {\cal L}_{\psi,{\rm mech}} & \equiv {\cal L}^{(1)}_{\psi} 
     = {\boldsymbol{\Omega}} \cdot {\boldsymbol{J}}_\psi\,,
     \label{eq_L_psi_1}
\end{align}
where $\psi$ is the Dirac 4-spinor and ${\bar \psi} = \psi^\dagger \gamma^t$ is its Dirac conjugate. The last term in Eq.~\eqref{eq_L_psi_1} is linear in the local angular momentum 
\begin{align}
    \boldsymbol{J}_\psi \equiv \boldsymbol{J}_\psi(\boldsymbol{x}) = \psi^\dagger(\boldsymbol{x}) {\hat{\boldsymbol{J}}} \psi(\boldsymbol{x})\,,
\end{align}
carried by the fermion field $\psi$. The $z$-component of the angular momentum operator, which enters the last term in Eq.~\eqref{eq_L_psi_1}, is given by the sum of the angular momentum operator and the spin operator, respectively:
\begin{align}
\hat{J}_{\psi,z}  
= -i\left(-y\partial_x + x\partial_y \right) + \frac{1}{2}
\Sigma^{xy}
\equiv -i\partial_\varphi + \frac{1}{2}
\begin{pmatrix}
\sigma_3 & 0 \\
0 & \sigma_3 
\end{pmatrix}
\,.
\label{eq_J}
\end{align}
In the last line, the $\Sigma^{xy}$ component of the spin matrix $\Sigma^{\mu\nu} = (i/2) [\gamma^\mu,\gamma^\nu]$ is written in the Dirac representation with $\sigma_3$ being a Pauli matrix.  

The Dirac Lagrangian~\eqref{eq_L_psi} describes a neutral massive spinor field in the co-rotating reference frame equipped with the metric~\eqref{eq_metric}. The Dirac Lagrangian in the laboratory frame~\eqref{eq_L_psi_0}, ${\mathcal L}_{\rm lab} \equiv {\cal L}(\Omega = 0)$, is given by the very same Eq.~\eqref{eq_L_psi} with angular frequency set to zero, $\Omega = 0$. 

It appears that the difference between the Lagrangians in the laboratory and co-rotating frames originates in the single term~\eqref{eq_L_psi_1} that includes the angular momentum~\eqref{eq_J}. The meaning of the term produced by the rotating environment is easy to estimate if one considers the Hamiltonian of the system~\eqref{eq_L_psi}:
\begin{align}
    {\hat H}_\psi = {\hat H}_{\psi,{\rm lab}} - {\boldsymbol{\Omega}} {\hat{\boldsymbol{J}}_\psi}\,.
    \label{eq_H_psi}
\end{align}
The rotating environment shifts the energy levels of the system via the mechanical coupling between the background angular velocity~$\boldsymbol{\Omega}$ and the total angular momentum of the system ${\boldsymbol{J}}_\psi$. The latter quantity comprises both the orbital and spin contributions. This natural type of mechanical coupling appears in classical mechanics~\cite{LL1} and thermodynamics~\cite{LL5}. 

For the reasons which will be clear shortly after, we call the last term in the Hamiltonian~\eqref{eq_H_psi} the ``mechanical-vortical coupling''. This term is responsible for various effects, including the standard classical mechanical phenomena exhibited by the rotating systems, such as the appearance of the Coriolis force and the centrifugal force~\cite{Matsuo2015}. The last statement is in no way trivial because the centrifugal force in mechanical systems is proportional to the square of the angular velocity, $\Omega^2$. In contrast, the coupling of the angular momentum of the system with the background angular velocity in the Hamiltonian~\eqref{eq_H_psi} has a linear nature. 

It is evident that the very same considerations also apply to quarks. The Dirac Lagrangian~\eqref{eq_L_psi} will then be modified to include the coupling to gluons via a covariant derivative as well as color and flavor indices of the Dirac spinors. The Hamiltonian for quarks will then be described by the general form given in Eq.~\eqref{eq_H_psi}, with the mentioned modifications of its terms.

\subsubsection{Gluons}

Gluon fields in the co-rotating frame are described by SU(3) Yang-Mills theory in the curved spacetime endowed with the metric~\eqref{eq_metric}. The Lagrangian in the Minkowski spacetime has the following form:
\begin{align}
    {\cal L}_G = - \frac{1}{4 g_{YM}^2} g^{\mu \nu} g^{\alpha \beta} F_{\mu \alpha}^a F_{\nu \beta}^a\,,
    \label{eq_L_G}
\end{align}
where $F_{\mu\nu}^a$ is the gluon field stress tensor and $g_{YM}$ is the Yang-Mills coupling constant. Substituting the metric~\eqref{eq_metric} into Eq.~\eqref{eq_L_G} allows us to decompose the Yang-Mills Lagrangian into three terms:
\begin{align}
    {\cal L}_G & = {\cal L}^{(0)}_{G} \hskip 3mm + {\cal L}^{(1)}_{G}  \hskip 6mm  + {\cal L}^{(2)}_{G} \nonumber\\
               & \equiv {\cal L}_{G,{\rm lab}} +  {\cal L}_{G,{\rm mech}} +  {\cal L}_{G,{\rm magn}}\,,
\label{eq_L_G_decomposition}
\end{align}
which, contrary to the fermionic Lagrangian~\eqref{eq_L_psi}, is a higher-order, quadratic polynomial of the angular frequency $\Omega$ with ${\cal L}^{(n)}_{G} \propto \Omega^n$ for $n=0,1,2$. Each term in Eq.~\eqref{eq_L_G_decomposition} has a certain physical meaning, which we discuss below. 

\vskip 2mm
\paragraph*{\bf (i) Laboratory frame term.}
The first term in Eq.~\eqref{eq_L_G_decomposition} is given by the Yang-Mills Lagrangian in the inertial (non-rotating) laboratory frame:
\begin{align}
    {\cal L}^{(0)}_G = - \frac{1}{4 g_{YM}^2} \eta^{\mu \nu} \eta^{\alpha \beta} F^a_{\mu \alpha} F^a_{\nu \beta}\,,
    \label{eq_L_G_0}
\end{align}
where $\eta^{\mu\nu} = {\rm diag}(+1,-1,-1,-1)$ is the flat Minkowski metric. The other two terms in Lagrangian~\eqref{eq_L_G_decomposition} encode the effects of the global rotation. 

\vskip 2mm
\paragraph*{\bf (ii) Mechanical coupling.}
The second contribution to Lagrangian~\eqref{eq_L_G_decomposition} is a linear term in the angular frequency $\Omega$. It has precisely the same generic form as the coupling for the fermions~\eqref{eq_L_psi_1}:
\begin{align}
    {\cal L}_{G,{\rm mech}} & \equiv {\cal L}^{(1)}_{G} = {\boldsymbol{\Omega}} \cdot {\boldsymbol{J}}_G\,,
\label{eq_L_G_1}
\end{align}
This term encodes the mechanical coupling between the background angular velocity $\boldsymbol{\Omega}$ and the angular momentum density of gluons, ${\boldsymbol{J}}_G$. The gluonic angular momentum density, taken in the non-rotating limit ($\Omega \to 0$), has the following components:
\begin{align}
    J_{G,i} & = \frac{1}{2} \varepsilon_{ijk} J^{jk}_G,
    \qquad
    J^{ij}_G = x^i T^{jt}_G - x^j T^{it}_G\,,
    \label{gl_ang_mom}
\end{align}
where
\begin{align}
    T^{\mu\nu}_G = - \frac{1}{g_{YM}^2} \left( F^{a,\mu\alpha} F^{a,\nu}_{\quad\, \alpha} - \frac{1}{4} \eta^{\mu\nu} F^{a,\alpha\beta} F^a_{\alpha\beta} \right)\,,
    \label{eq_T_munu} 
\end{align}
is the gluonic stress-energy tensor in the Minkowski spacetime. The rising and lowering of indices in Eq.~\eqref{eq_T_munu} are performed using the Minkowski metric $\eta^{\mu\nu}$. 

The gluonic angular momentum, given in Eqs.~\eqref{gl_ang_mom} and  \eqref{eq_T_munu}, in laboratory frame has the following familiar form:
\begin{align}
    {\boldsymbol{J}}_G = \frac{1}{g_{YM}^2} {\boldsymbol{r}} \times ({\boldsymbol{E}}^a \times {\boldsymbol{B}}^a)\,,
\label{eq_J_G}
\end{align}
where ${\boldsymbol{S}} \equiv \frac{1}{g_{YM}^2} ({\boldsymbol{E}}^a \times {\boldsymbol{B}}^a)$ is the Poynting vector which represents the directional energy flux for the gluon field. Notice that both chromoelectric, $E^a_i = F^a_{ti}$, and chromomagnetic, $B^a_i = - (1/2) \varepsilon_{ijk} F^{a,jk}$, fields contribute to the gluon angular momentum~\eqref{eq_J_G}.

\vskip 2mm
\paragraph*{\bf (iii) Magnetic coupling.} The gluon Lagrangian~\eqref{eq_L_G_decomposition} also has a term quadratic in the angular frequency $\Omega$:
\begin{align}
    {\cal L}_{G,{\rm magn}} \equiv {\cal L}^{(2)}_{G} = 
    \frac{1}{2 g_{YM}^2} \Bigl[ \Omega^2 ({\bs B}^a \cdot {\bs r}_\perp)^2 +  {r^2_\perp} ({\bs B}^a \cdot {\bs \Omega})^2 \Bigr]\,,
\label{eq_L_G_2}
\end{align}
where ${\bs r}_\perp = {\boldsymbol{r}} - z {\bf e}_z \equiv {\bf e}_z \times ({\bf r} \times {\bf e}_z)$ is a vector which points, along a normal, from the axis of rotation to the point $\boldsymbol{r} = (x,y,z)$. The quadratic term~\eqref{eq_L_G_2} couples the chromomagnetic field ${\boldsymbol{B}}^a$ to the background angular velocity $\boldsymbol{\Omega}$. Therefore, we call this term a ``magnetovortical coupling''.

One could naively suspect that the magnetovortical term~\eqref{eq_L_G_2}, being quadratic in $\Omega$, is responsible for the net centrifugal force that occurs in a rotating gluon gas. However, while this term certainly contributes to the mentioned force, it does not lie in its origin. To see this fact, one can note that the centrifugal force does exist for fermions that do not possess the quadratic in $\Omega$ coupling in the fermionic Lagrangian~\eqref{eq_L_psi}.

The Lagrangian of gluons in rotating frame is written in terms of rotating gluon fields, and it is a quadratic function in the angular velocity~\eqref{eq_L_G_decomposition}, highlighting the magnetovortical coupling. However, the Hamiltonian of the rotating gluons has the usual linear form
\begin{align}
    {\hat H}_G = {\hat H}_{G,{\rm lab}} - {\boldsymbol{\Omega}} {\hat{\boldsymbol{J}}_G}\,,
    \label{eq_H_G}
\end{align}
as it was explicitly shown in Refs.~\cite{Yang:2023vsw, Wang:2025mmv},
and the static system in rotating frame corresponds to rotating system in the laboratory frame.
The key distinction between gluons and quarks is that the gluons are vector fields, and they transform when we pass from one to another reference frame.
As the result,  Eqs.~\eqref{eq_L_G_decomposition} and~\eqref{eq_H_G} have different structure, and the magnetic coupling arises explicitly only in the rotating frame. 
It implies that   some degrees of freedom in the rest frame might play a role of the magnetic coupling in rotating frame, but it is a non-trivial task to separate them explicitly in the laboratory frame.

It is also worth noticing that the canonical analyses of the thermodynamics of rotating systems are conventionally formulated in the rotating reference frame, in which the temperature and other thermodynamic parameters (such as the chemical potentials) are specified~\cite{LL5, LL9}. A calculation in the laboratory reference frame lies outside of the scope of the present paper since here we concentrate on the conventional analysis in the corotating reference frame. A discussion of the relation between calculations in the two frames in the scope of the lattice gluodynamics can be found in Refs.~\cite{Yang:2023vsw, Wang:2025mmv}.

\subsection{Exotic vortical properties of gluonic matter}

Our first-principle numerical simulations~\cite{Braguta:2023yjn, Braguta:2023kwl, Braguta:2023tqz, Braguta:2023qex} revealed that hot gluonic matter has exotic vortical properties in the phenomenologically important window of temperatures that ranges from slightly lower than the deconfining transition at $T = T_c$ up to supervortical temperature $T_{s} \simeq 1.5 T_c$. The most striking feature of the gluon plasma at these temperatures is that it possesses a negative moment of inertia~\cite{Braguta:2023tqz}. 

In a mechanical sense, the microscopic mechanism of this exotic effect has been proposed to originate from the negative Barnett effect, which implies the negative value of the spin-vortical coupling in rotating hot gluon matter.~\cite{Braguta:2023tqz} The negativity of the coupling has an intuitively clear meaning, indicating that in the rotating gluonic fluid, the total spin polarization of vortical gluons and their total orbital momentum should be oriented in opposite directions. Since the orbital momentum points in the direction of the angular velocity, the total angular momentum ${\boldsymbol{J}}$ stored in the vortical plasma becomes directed opposite to its angular velocity $\boldsymbol{\Omega}$. The latter fact implies the negativity of the moment of inertia, $I < 0$, which is nothing but the coefficient of proportionality between the angular momentum and the angular velocity of the body, ${\boldsymbol J} = I {\boldsymbol \Omega}$. 

Initially, the negative moment of inertia has been associated with the instability of the host gluon plasma with respect to the vortical motion~\cite{Braguta:2023yjn}. However, it later became clear that the plasma can rotate in the thermodynamic equilibrium while the negative moment of inertia can qualitatively be explained by the negative Barnett effect~\cite{Braguta:2023tqz}. The latter property is a particular feature of the strongly interacting vector particles, gluons. 

The moment of inertia for a generic statistical system is determined via its free energy~$F$ in the co-rotating reference frame~\cite{LL5}:
\begin{align}
    I(T) = \frac{J(T,\Omega)}{\Omega} {\biggl|}_{\Omega \to 0} \equiv - \frac{1}{\Omega} {\left( \frac{\partial F(T,\Omega)}{\partial \Omega} \right)}_{T}{\biggl|}_{\Omega \to 0}\,.
\label{eq_I_F}
\end{align}
For gluodynamics, the free energy $F = F_G$ will be defined in the next section.

The moment of inertia of gluons can be represented as a sum of two parts~\cite{Braguta:2023tqz}:
\begin{align}
	I_G = \frac{1}{T} \aavr{ \bigl({\bs n} \cdot {\bs J}^{\rm tot}_G \bigr)^2}_T 
    + \int d^3 x \Bigl[\aavr{({\bs B}^a \cdot {\bs r}_\perp)^2}_T + \aavr{({\bs B}^a \cdot {\bs n})^2}_T {r_\perp^2} \Bigr]\,,
    \label{eq_I_G}
\end{align}
where ${\boldsymbol{n}} = {\boldsymbol{\Omega}/\Omega}$ is the axis of rotation, $\aavr{{\mathcal O}}_T = \left\langle {\mathcal O} \right\rangle_T - \left\langle {\mathcal O} \right\rangle_{T=0}$ is the thermal part of the expectation value of an operator $\mathcal O$, and ${\boldsymbol{J}}^{\rm tot}_G$ is the total gluonic angular momentum~\eqref{eq_J_G} integrated over the volume of the system (a rigorous definition is given in \cite{Braguta:2023tqz}).

The first term in Eq.~\eqref{eq_I_G} appears due to the linear coupling~\eqref{eq_L_G_1} of the angular gluonic degrees of freedom to the angular velocity. This term has a form of a susceptibility which represents a conventional mechanical contribution to the moment of inertia. As noted in Ref.~\cite{Braguta:2023tqz}, the mechanical term gives always a non-negative contribution to the moment of inertia because $\aavr{\bigl({\bs n} \cdot {\bs J}^{\rm tot}_G \bigr)^2}_T \geqslant 0$.

The presence of the last term in Eq.~\eqref{eq_I_G} is the consequence of the quadratic coupling~\eqref{eq_L_G_2} of gluons to the angular velocity, which appears in the gluonic Lagrangian in the rotating reference frame~\eqref{eq_L_G_decomposition}. While the linear coupling is commonly encountered across a wide range of statistical systems, the quadratic coupling has an exotic nature, which can be traced back to Ref.~\cite{Yamamoto:2013zwa}.

At the level of quantum chromodynamics, the negativity of the moment of inertia can be associated with the thermal melting of the chromomagnetic condensate~\cite{Braguta:2023yjn} which provides a negative contribution to the second term in Eq.~\eqref{eq_I_G}. The gluon condensate is a nonperturbative quantity proportional to the expectation value of the gluon field strength squared, $\avr{\Tr F_{\mu\nu} F^{\mu\nu}} > 0$. As we have already mentioned, the magnetovortical term~\eqref{eq_L_G_2} couples the bilinear of the chromomagnetic field to (a square of) the angular velocity and, thus, provides a contribution of the chromomagnetic condensate to the moment of inertia. The first-principle numerical calculations in Ref.~\cite{Braguta:2023tqz} demonstrated that this contribution takes a negative value in a relevant temperature window. Thus, the $I < 0$ effect has been attributed to the thermal melting of the gluon condensate.\footnote{The gluon condensate is a gauge-invariant quantity. Notice that at zero temperature, it also possesses the Lorentz invariance, implying that its division into chromomagnetic and chromoelectric components has no Lorentz-invariant sense. Indeed, the condensate is described by the Lorentz invariant quantity $\Tr F_{\mu\nu} F^{\mu\nu} \equiv  (\boldsymbol{B}^a)^2 -  (\boldsymbol{E}^a)^2$ and not by the electric field $\boldsymbol{E}^a$ or the magnetic field $\boldsymbol{B}^a$ separately. Still, we use the generally accepted term ``chromomagnetic condensate'' because, in the nonperturbative ground state, one has $\avr{\Tr F_{\mu\nu} F^{\mu\nu}} > 0$, which is interpreted as the prevalence of the chromomagnetic condensate over the chromoelectric one~\cite{Savvidy:1977as}.} 
The present paper aims to elucidate further the expected crucial role of the magnetovortical term~\eqref{eq_L_G_2} in the exotic vortical properties of hot gluon matter.

\section{Lattice simulation of rotating gluodynamics}\label{sec_Formulation}

\subsection{Theoretical background}

We study the rotating matter of hot gluons using first-principle numerical simulations of the lattice Yang-Mills theory.\footnote{We follow the approach of Refs.~\cite{Braguta:2021jgn, Braguta:2022str} to which we refer an interested reader for technical details.}
The matter resides in thermodynamic equilibrium in the non-inertial co-rotating reference frame, which is described by metric~\eqref{eq_metric}. The partition function in the Euclidean spacetime, suitable for Monte Carlo simulations can be written as a path integral over the gluon fields
\begin{equation}\label{eq_Z} 
   Z \equiv e^{- F_G/T} = \int \!D A\, e^{-S_G[A]}\,. 
\end{equation}
where $F_G$ is the free energy of gluons. The $S_G$ in Eq.~\eqref{eq_Z} is the Euclidean action of rotating Yang-Mills theory:
\begin{equation} \label{eq_S_E}
S_G = \frac{1}{4 g_{YM}^2} \int d^{4} x\, \sqrt{g_E}\,  g^{\mu \nu}_E g^{\alpha \beta}_E F_{\mu \alpha}^{a} F_{\nu \beta}^{a}\,,
\end{equation}
where $F_{\mu\nu}^a$ is the field-strength tensor of the gluon field. The Euclidean metric $g_{\mu \nu}^E$ is related to the Minkowski metric~\eqref{eq_metric} via the Wick rotation $t \to -i \tau$. The imaginary-time direction $\tau$ is compactified to a cirle of the length of the inverse temperature $1/T_0$ with periodic boundary conditions for the gluon fields. The explicit form of the Euclidean action~\eqref{eq_S_E} is as follows:
\begin{multline}\label{eq_S_E_continuum}
	S_{G} = \frac{1}{2 g_{YM}^2 } \int\! d^{4}x \ \Big[
    F^a_{x \tau} F^a_{x \tau} + F^a_{y \tau} F^a_{y \tau} + F^a_{z \tau} F^a_{z \tau} + {} \\
    + \left(1 - x^2 \Omega^2\right) F^a_{y z} F^a_{y z} 
    + \left(1 - y^2 \Omega^2\right) F^a_{x z} F^a_{x z} 
    + \left(1 - r^2 \Omega^2\right) F^a_{x y} F^a_{x y} - {} \\
    - 2 y i \Omega \left(F^a_{x y} F^a_{y \tau} + F^a_{x z} F^a_{z \tau}\right) 
    + 2 x i \Omega \left(F^a_{y x} F^a_{x \tau} + F^a_{y z} F^a_{z \tau}\right)
    - 2 x y \Omega^2 F^a_{x z} F^a_{z y} 
    \Big]\,.
\end{multline}
For later use, we rewrite this action in cylindrical coordinates $d x^{\mu} = (d \tau, dr, r d \varphi, d z)$ where it has a quite simple form:
\begin{multline} 
\label{eq_S_E_continuum_c}
    S = \frac {1} {2 g_{YM}^2 } \int d^4x \Big[ 
F^a_{r \tau} F^a_{r \tau} + F^a_{\hat \varphi \tau} F^a_{\hat \varphi\tau } + F^a_{z \tau} F^a_{z \tau} 
    {} + F^a_{rz} F^a_{rz} + {} \\
    + \left(1-r^2\Omega^2 \right) \left(F^a_{\hat \varphi z} F^a_{\hat \varphi z} + F^a_{r \hat \varphi} F^a_{r \hat \varphi} \right) 
{} + 2 r i \Omega \left( F^a_{\hat \varphi r} F^a_{r \tau} + F^a_{\hat \varphi z} F^a_{z \tau} \right) \Big]\,, 
\end{multline}
where $\hat \varphi$ denotes the polar direction with $d \hat \varphi = r d \varphi$.

Notice that the thermodynamic equilibrium of a system of particles subjected to an external static gravitational field has, in general, an inhomogeneous temperature that encodes the redshift of the thermal wavelength by a gravitational force. The local temperature is given by the Ehrenfest–Tolman (TE) law: $T(\bs r) \sqrt {g_{00}} = const$~\cite{Tolman:1930zza, Tolman:1930ona}. For a rigidly rotating system with the metric~(\ref{eq_metric}), the TE law gives $T(r) \sqrt {1- r^2 \Omega^2} = T_0$~\cite{Braguta:2021jgn},
where $T_0 = T(r=0)$ corresponds to the temperature at the axis of rotation, which is determined by the extension $1/T_0$ of the lattice in the temporal direction. For notational simplicity, we denote hereafter by $T$ the value of temperature precisely at the rotation axis, where the metric-induced effects vanish unless stated otherwise.

\subsection{Lattice set up}\label{sec_Formulation_setup}

In this work, we discretize rotating terms in the action~\eqref{eq_S_E_continuum} following Refs.~\cite{Yamamoto:2013zwa, Braguta:2021jgn} and use the tree-level improved Symanzik gauge action for the terms without rotation~\cite{Curci:1983an, Luscher:1985zq}. The explicit lattice expression for action~\eqref{eq_S_E_continuum} can be found in Ref.~\cite{Braguta:2023yjn}.

We perform the simulations on the lattices with the spacetime geometry $N_t \times N_z \times N_s^2$ ($N_x = N_y = N_s$). The rotational axis is located at the center of the $xy$ plane. In our simulations, the transverse spatial extension of the lattice $N_s$ is an odd number, so the rotation axis passes through the sites of our lattice. It is worth mentioning that the temporal extension $N_t$ is related to the temperature $T$ as $T=T_0=1/(aN_t)$, where $a$ is the lattice spacing. The size of the lattice along the axis of rotation $N_z$ is arbitrary, and one can take an infinite volume limit in $z$-direction. On the contrary, one has to restrict the size of the lattice in the transverse plane so that the velocity at any point of the considered volume is smaller than the speed of light. Taking into account the rectangular geometry of the lattices, the velocity $v$ at the middle point of the boundary plane is limited as $v = \Omega R < 1/\sqrt{2}$, where $R = a(N_s - 1)/2$.

Boundary conditions are particularly important in the simulations. In our study, we impose periodic boundary conditions in $\tau$- and $z$-directions. What concerns $x$- and $y$-directions, one can impose open, periodic, and Dirichlet boundary conditions. These boundary conditions were studied in paper~\cite{Braguta:2021jgn}. It was found that though there is some dependence on boundary conditions, this dependence doesn't play a significant role for bulk observables because of screening. Nevertheless, in order to reduce the influence of boundary on our results, we perform simulations on the lattices with large $N_s$: $N_s/N_t \gtrsim 24$. In this paper mostly, we are going to use open boundary conditions~(OBC). In addition to studying the systematic uncertainty of our results, in some cases, periodic boundary conditions~(PBC) will be employed. 

Action~(\ref{eq_S_E_continuum}), as well as its lattice version, are complex-valued functions. As a result, Monte Carlo methods cannot be applied for the direct calculation of the partition function of this system because of the sign problem. In order to overcome this difficulty, we carry out a lattice study of rotating gluodynamics at imaginary values of angular frequency~$\Omega_I$,
\begin{align}
    \Omega_I = -i \Omega\,.
\end{align}
We also introduce the imaginary velocity at the boundary~$v_I$:
\begin{align}
    v_I = \Omega_I R \equiv \Omega_I a \frac{N_s - 1}{2}\,.
    \label{eq_v_I}
\end{align}
The results obtained in this way can be analytically continued to real values of angular frequency. Note that this continuation works only for the bounded systems subjected to the causality condition~\cite{Ambrus:2023bid}, which can be formulated for the imaginary angular frequencies on the square-shaped cylinder as $v_I < 1/\sqrt{2}$. The causality condition is satisfied in our lattice simulations.

\subsection{Phase transition and Polyakov loop}

In the absence of global vorticity, Yang-Mills theory has two homogeneous phases: the low-temperature confinement phase and the high-temperature deconfinement phase. In order to distinguish these phases, the expectation value of the Polyakov loop is commonly used
\begin{align} \label{eq_L_continuum}
    L({\bs r}) = {\mathrm {Tr}}\, {\mathcal P} \exp \biggl(\oint_{0}^{1/T} d \tau A_4( {\tau},{\bs r}) \biggr) \,,
\end{align}
where ${\mathcal P}$ is the path-ordering operator. On the lattice, it has the following definition:
\begin{equation}\label{eq_polyakov_loop}
    L(\bs r) = \Tr \left[ \prod_{\tau = 0}^{N_t - 1} U_4(\tau, \bs r) \right]\,,
\end{equation}
where $U_4(\tau, \bs r)$ is the link variable in the temporal direction.

The Polyakov loop can be parameterized by the free energy of static quark $F_{Q}$ as $\langle L \rangle = e^{-F_Q/T}$. In the confinement phase, infinite energy is needed to create a free quark, whereas, in the deconfinement, the free energy of a static quark is finite. As a result, for static hot gluon matter, the Polyakov loop takes a zero value in the confinement phase, and this quantity is globally nonzero in the deconfinement phase. 

The analytical form of the action~\eqref{eq_S_E_continuum} implies that rotating gluodynamics is homogeneous in $z$-direction and inhomogeneous in the transverse directions. Therefore, the homogeneity of the gluon configurations is lost in the vortical plasmas. However, we believe that the local expectation value of the Polyakov loop remains an order parameter also for rotating gluodynamics, and this local order parameter can be used to distinguish spatially separated confinement and deconfinement phases.  

To study the spatial distribution of the Polyakov loop, we introduce the local Polyakov loop in $x,y$-plane as
\begin{equation}\label{eq_polyakov_loop_xy}
    L(x,y) = \frac{1}{N_z} \sum_{z=0}^{N_z-1} L(x,y,z)\,,
\end{equation}
where $L(x, y, z) = L(\bs r)$ is defined by Eq.~\eqref{eq_L_continuum} and calculate its ensemble average. Using the local Polyakov loop~\eqref{eq_polyakov_loop_xy}, one can introduce local Polyakov loop susceptibility, 
\begin{equation}\label{eq_usceptibility}
\chi_L(r) = \langle |L(r)|^2 \rangle  - \langle |L(r)|\rangle^2\,,
\end{equation}
which can also be employed to detect the spatial confinement/deconfinement transition.

Lattice calculations are performed at finite $N_z$, which leads to high uncertainty in the determination of the position of the confinement/deconfinement transition. To reduce the associated statistical error, we calculated the mentioned quantities within a thin cylinder $(r - \delta r/2, r + \delta r/2)$. We justified our approach by demonstrating numerically that the finiteness of $\delta r$ brings only a minor systematic error to the estimation of the critical temperature, as we demonstrate below.

In this work, we focus on the confinement–deconfinement phase transition, characterized via the Polyakov loop and its susceptibility. The deconfining transition point is conventionally identified either by the inflection point of the Polyakov loop ---marking its change from near-zero to finite values--- or by the peak of the susceptibility of the Polyakov loop. These points coincide in the thermodynamic limit for a genuine phase transition. In pure SU(3) Yang–Mills theory, where the transition is weakly first-order, the location of the transition point can be determined directly from the bare (unrenormalized) Polyakov loop and its susceptibility.

\section{Mixed inhomogeneous phase in rotating gluon plasma} \label{sec_Coexistence}
\subsection{Observation of mixed inhomogeneous phase}

\begin{figure*}
    \includegraphics[width = 1.0\linewidth]{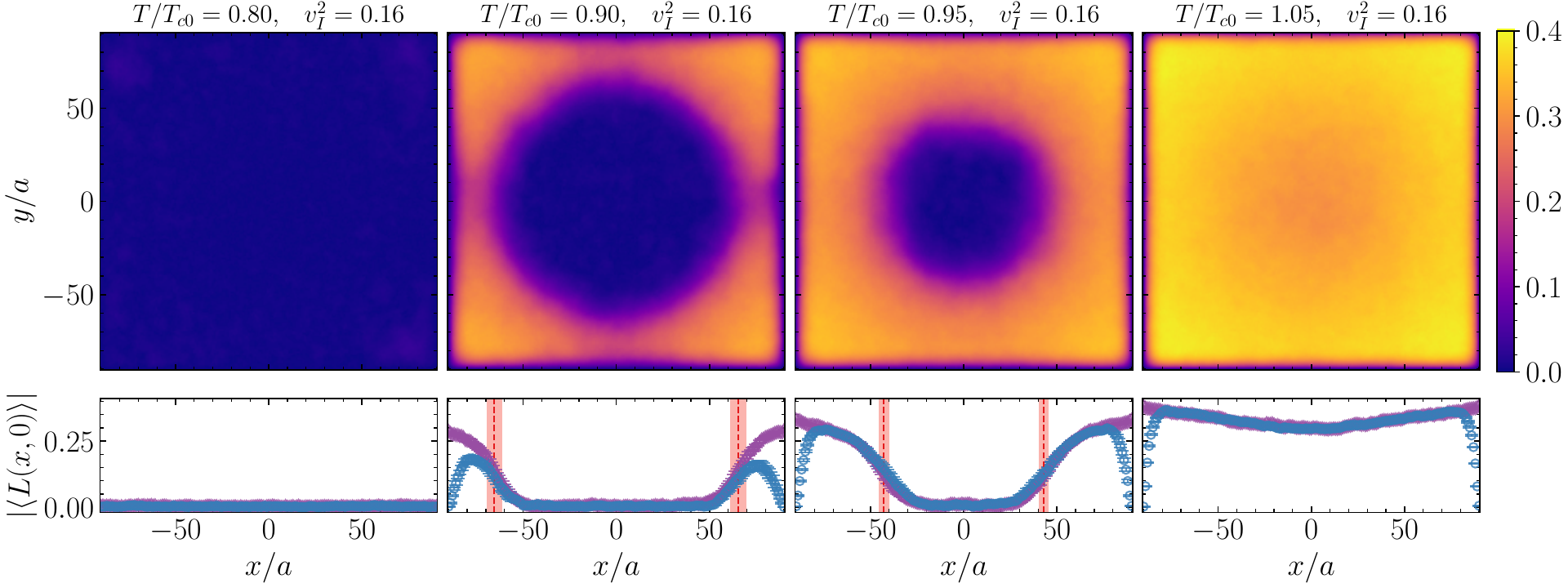}
    \caption{(top) The distribution of the local Polyakov loop in $x,y$-plane for lattice of size $5\times 30\times 181^2$ with open boundary conditions at the fixed imaginary velocity at the boundary $v_I^2 \equiv (\Omega_I R)^2 = 0.16$ and different on-axis temperatures $T/T_{c0}= 0.80,\, 0.90,\, 0.95,\, 1.05$.
    (bottom) The Polyakov loop at the $x$ axis $(y=0)$. The vertical lines mark the phase boundaries with shaded uncertainties. The violet (blue) data points correspond to periodic (open) boundary conditions.
    Movies showing the evolution of inhomogeneity as temperature rises are available as ancillary files (see Supplementary material).
    }
    \label{fig_imshowPLxy-v4}
\end{figure*}

Taking into account the explicit coordinate dependence of the action~\eqref{eq_S_E_continuum}, one can expect that various local observables are inhomogeneous quantities in the transverse $x,y$-plane, which is normal to the rotation axis. In this subsection, we study the spatial distribution of the Polyakov loop in the transverse plane following the first study of the inhomogeneous phases in vortical gluodynamics~\cite{Braguta:2023iyx}. In this section, we present an improved and detailed analysis of the results obtained {earlier} in Ref.~\cite{Braguta:2023iyx}.

We fix the imaginary boundary velocity at the value $v_I^2 = (\Omega_I R)^2 = 0.4$ and vary the on-axis temperature $T$ of our system. The distributions of the local Polyakov loop~\eqref{eq_polyakov_loop_xy} are shown in Fig.~\ref{fig_imshowPLxy-v4}, where the temperature is given in terms of the critical temperature $T_{c0}$ of a non-rotating gluon matter. The results presented in this subsection were obtained on the lattice $5\times 30 \times 181^2$. Notice that in this subsection, we only average the Polyakov loop over the $z$-direction.

At low temperatures (see the left panel in Fig.~\ref{fig_imshowPLxy-v4} with $T/T_{c0} = 0.80$), the system is in the confinement phase, and the local Polyakov loop fluctuates close to zero in the whole space. If we heat the system, it remains in this state up to a certain temperature when an inhomogeneous two-phase structure appears. Then, imaginary rotation results in the deconfinement phase at the periphery of the system, while the region near the rotation axis stays in the confinement phase. As the (on-axis) temperature increases, the radius of the inner region shrinks (see two central panels in Fig.~\ref{fig_imshowPLxy-v4} with $T/T_{c0} = 0.90,\,0.95$). Finally, at the temperature above the confinement/deconfinement transition of non-rotating gluodynamics, the whole volume is in the deconfinement phase (see the right panel in Fig.~\ref{fig_imshowPLxy-v4} with $T/T_{c0} = 1.05$).

\begin{figure*}
    \includegraphics[width = 1.0\linewidth]{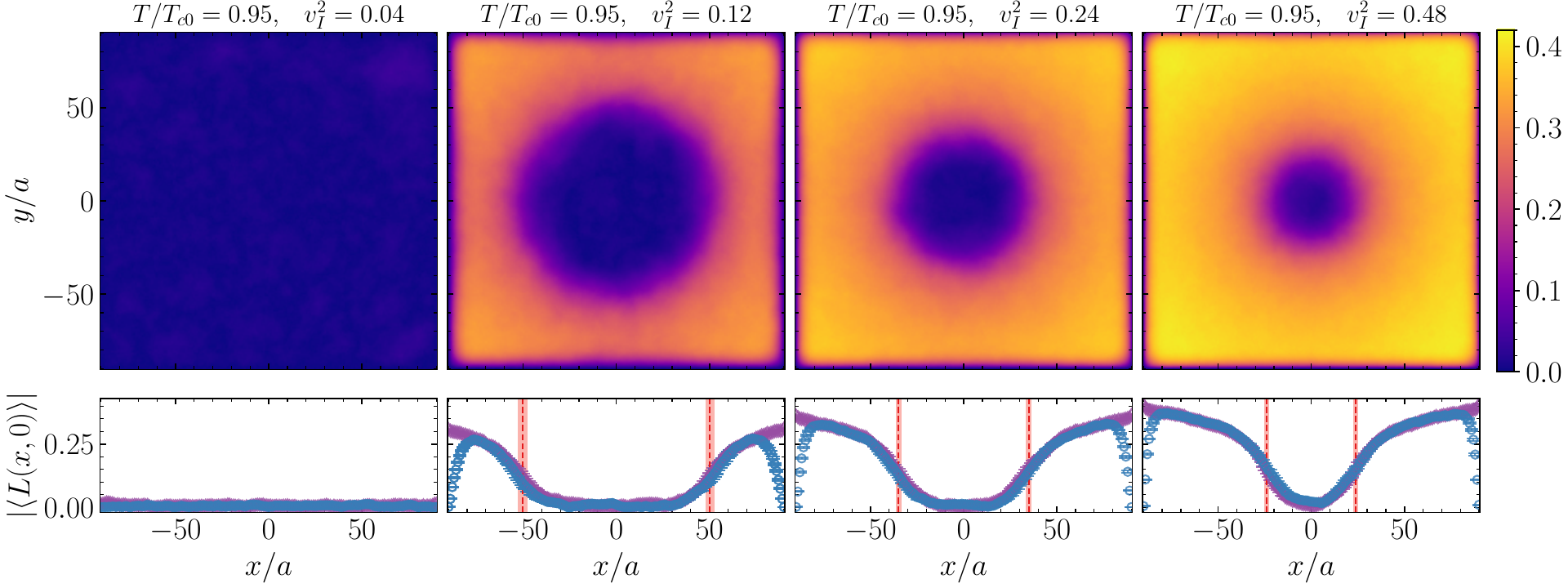}
    \caption{The same as in Fig.~\ref{fig_imshowPLxy-v4} but for the temperature $T/T_{c0} = 0.95$ and different imaginary velocities at the boundary, $v_I^2 \equiv (\Omega_I R)^2 = 0.04,\, 0.12,\, 0.24,\, 0.48$. Movies on the phase evolution with increasing angular velocity are available as ancillary files (see Supplementary material).
    }
    \label{fig_imshowPLxy-T095}
\end{figure*}

In our second numerical calculation, we fix the on-axis temperature at the value $T/T_{c0}=0.95$, which is slightly below the critical temperature of the non-rotating system, and vary the imaginary boundary velocity~$v_I$. The results are presented in Fig.~\ref{fig_imshowPLxy-T095}. From these figures, one sees that for a slow rotation, $v_I^2 = 0.04$, the whole system is in the confinement phase. For larger velocities $v_I^2=0.12,\, 0.24$, there appears to be a mixed inhomogeneous phase with deconfinement at the periphery and confinement in the center. This picture also holds for the largest imaginary velocity $v_I^2 = 0.48$, which is quite close to the causality limit $v_I^2 < 0.5$. Interestingly, the system in the center stays in the confinement phase even for such a fast rotation.

To summarize, our results indicate that the imaginary rotation of gluodynamics gives rise to the appearance of a mixed-phase state. In this state, the confinement phase is located in the center region, whereas the deconfinement phase is located in the periphery of the studied volume. For imaginary rotation, this mixed-phase state takes place for on-axis temperatures smaller than the critical temperature of the non-rotating system $T_{c0}$. For larger temperatures, $T>T_{c0}$, the mixed-phase state does not emerge, and the whole imaginary-rotating system resides in the deconfinement phase with a minor dependence of the local Polyakov loop on the radius (see the right panel in Fig.~\ref{fig_imshowPLxy-v4}).

The lower rows of Figs.~\ref{fig_imshowPLxy-v4} and \ref{fig_imshowPLxy-T095} show the dependence of our results on boundary conditions. These plots show that boundary conditions affect only small regions adjacent to boundaries. The results for PBC and OBC agree in bulk. The movies showing the evolution of the Polyakov loop distribution due to the variation of simulation parameters for both PBC and OBC are attached as ancillary files to this article (see Supplementary material).

We mention that our results do not depend on the size of the lattice in the $z$ direction. Notice also that as one can see from Fig.~\ref{fig_imshowPLxy-v4}, the regions with different phases are well spatially separated, and the transition region between these phases has a regular circle form, despite the square geometry of the lattice. Thus, cylindrical symmetry is restored in the bulk of our system for our lattice parameters. The restoration of the rotation symmetry is a natural consequence of the property that we work close to the continuum limit of the lattice Yang-Mills theory.

It is important to discuss the width of the transition region between phases. In our study, we have found that this width can be estimated as $\sim 3-5$~fm, and it weakly depends on the parameters of the simulations (see Appendix~\ref{app_width}). So, one can conclude that in order to observe the mixed inhomogeneous phase in rotating gluon plasma, one needs lattices of rather large sizes in the transverse $xy$ directions.

\subsection{Study of local critical temperature}\label{sec_Coexistance-localTc}

The mixed-phase picture assumes that the phase transition at different points of the rotating system occurs at different values of the (on-axis) temperature. To characterize this phenomenon quantitatively, we introduce the local (pseudo)critical temperature $T_c(r)$. This quantity is defined as the temperature on the rotation axis, for which the system,  with angular velocity $\Omega_I$, undergoes the confinement/deconfinement phase transition at a distance $r$. This function determines the position of the transition region. Indeed, for a given temperature $T = T_c(r)$, one has the confinement phase at distances smaller than~$r$ and the deconfinement phase at larger distances.

{In order to find the local critical} temperature, we calculate the expectation value of the Polyakov loop $L(r)$ and its susceptibility $\chi_L(r)$. The critical temperature $T_c(r)$ can be found from the peak position in the susceptibility $\chi_L(r)$. Notice, however, that to reach acceptable accuracy, one needs either lattices with large $N_z$ extent or quite large statistics. As was mentioned above, to overcome this problem, we average the Polyakov loop within a thin cylinder of the width $\delta r$. In addition, to reduce the influence of boundary conditions, we discard $\delta b$ layers adjacent to the boundary and do not account for them in the calculation. Technical details of the fitting procedure can be found in Appendix~\ref{app_Fit}.

\begin{figure}[t]
    \centering
    \includegraphics[width = 0.60\linewidth]{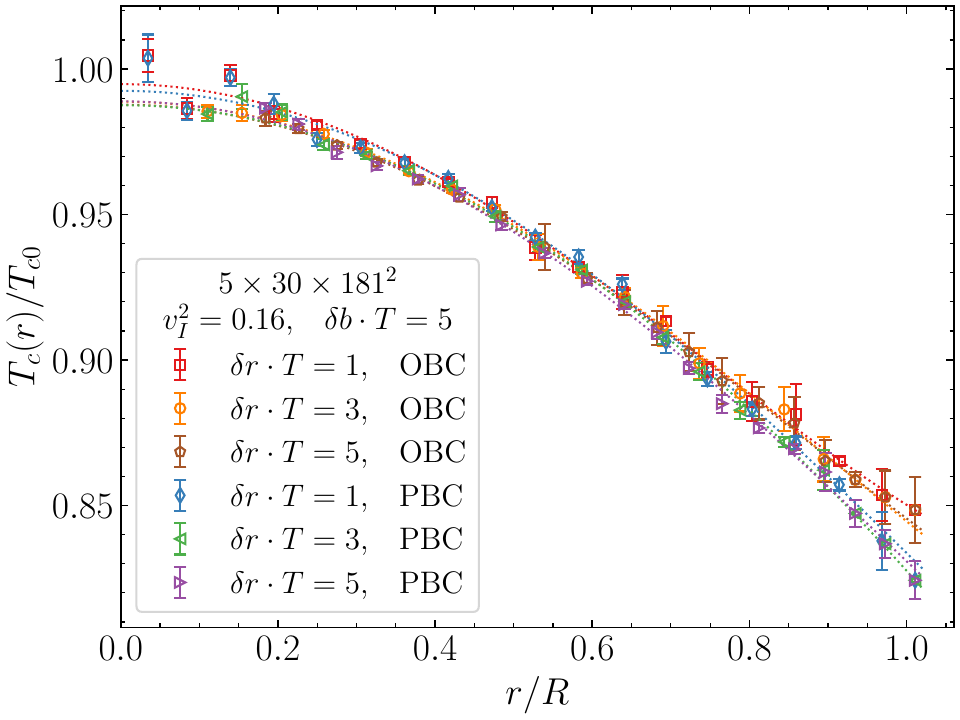}
    \caption{The local critical temperature (shown in units of the critical temperature $T_{c0}$ of nonrotating Yang-Mills theory) as a function of radial distance for the periodic (PBC) and open (OBC) boundary conditions, several values of the averaging width $\delta r \cdot T = 1,\,3,\,5$ and {the width of the ignored boundary layer} $\delta b\cdot T=5$ at the fixed velocity of rotation $v_I^2 = (\Omega_I R)^2 = 0.16$. The dotted lines show the fit of the data points by function~\eqref{eq_fit_4}.
    }
    \label{fig_tcloc-compBC-compW-nt5_1}
\end{figure}

In order to study how our results depend on the simulation parameters, we calculated the critical temperature $T_c(r)$ for the fixed value of the imaginary velocity $v_I^2=(\Omega R)^2=0.16$. The simulation was carried out on the lattice $5\times 30\times 181^2$ with periodic/open boundary conditions, {the width of ignored boundary layer} $\delta b \cdot T = 5$, and the following widths of the circular layer: $\delta r \cdot T = 1,\,3,\,5$. The results are presented in Fig.~\ref{fig_tcloc-compBC-compW-nt5_1}.

One can see that the results for different $\delta r$ are in good agreement with each type of boundary condition. The smaller width of the averaging layer $\delta r$ allows us to compute the local critical temperature closer to the rotation axis. However, as can be anticipated, the uncertainty of the calculation with smaller $\delta r$ is larger.

Although the results shown in Fig.~\ref{fig_tcloc-compBC-compW-nt5_1} were obtained with relatively large width of the ignored boundary layer $\delta b \cdot T = 5$, one can still distinguish some difference between the results for open and periodic boundary conditions at $r/R \sim 1$. However, this difference is quite small, i.e., we can state that boundary effects are under control in our simulations. We analyzed the results for several thicknesses of the boundary layer $\delta b$ and found that for $\delta b \cdot T \gtrsim 3$, boundary effects do not play a significant role. 

To fit the data for the dependence of the critical temperature on the distance $r$ from the axis or rotation, we used the following quartic formula 
\begin{equation}\label{eq_fit_4} 
    \frac{T_c(r)}{T_{c0}} = C_0 - C_2 \left( \frac{r}{R}\right)^2 + C_4 \left( \frac{r}{R}\right)^4\,.
\end{equation}
We found that the quality of the fit normalized by the number of the degrees of freedom (n.d.f.) used in the fit are $\chi^2/{\rm n.d.f.} \sim 1$ for $\delta r \cdot T \geq 3$ and $\chi^2/{\rm n.d.f.} \sim 2-3$ for $\delta r \cdot T = 1$, {indicating that the fit~\eqref{eq_fit_4} describes our data reasonably well}. The results for the best fit are shown in Fig.~\ref{fig_tcloc-compBC-compW-nt5_1} by dotted lines. 

In addition, we tried a quadratic fit, i.e., by setting the quartic constant to zero, $C_4=0$, to describe our data. This approximation appears to be inconsistent with our results if the data is taken in the whole range in $r$. However, the parabolic fit can still be used if one restricts the region {of the fit} by $r/R \lesssim 0.5$, as it was done in Ref.~\cite{Braguta:2023iyx}. In this case, the fit quality is good $\chi^2/{\rm n.d.f.}\sim 1-3$, and the results of the $C_0, C_2$ are in agreement with that obtained through the fitting by the formula (\ref{eq_fit_4}) in the whole available region. 

Further, we consider the dependence of the coefficients $C_i$, $i=0,2,4$ on the lattice spacing $a$ and transverse size of the lattice used in this simulation. We found that the $C_0$ and $C_2$ coefficients depend weakly on these parameters for the OBC and PBC, while the $C_4$ shows certain dependence on particularities of the lattice setup. In Appendix~\ref{app_NsNt}, we present the relevant details and the results of this analysis. Since our results weakly depend on the boundary conditions, hereafter, we are going to use open boundary conditions, which are more physically motivated than the periodic ones.

\begin{figure}
    \centering
    \includegraphics[width = 0.60\linewidth]{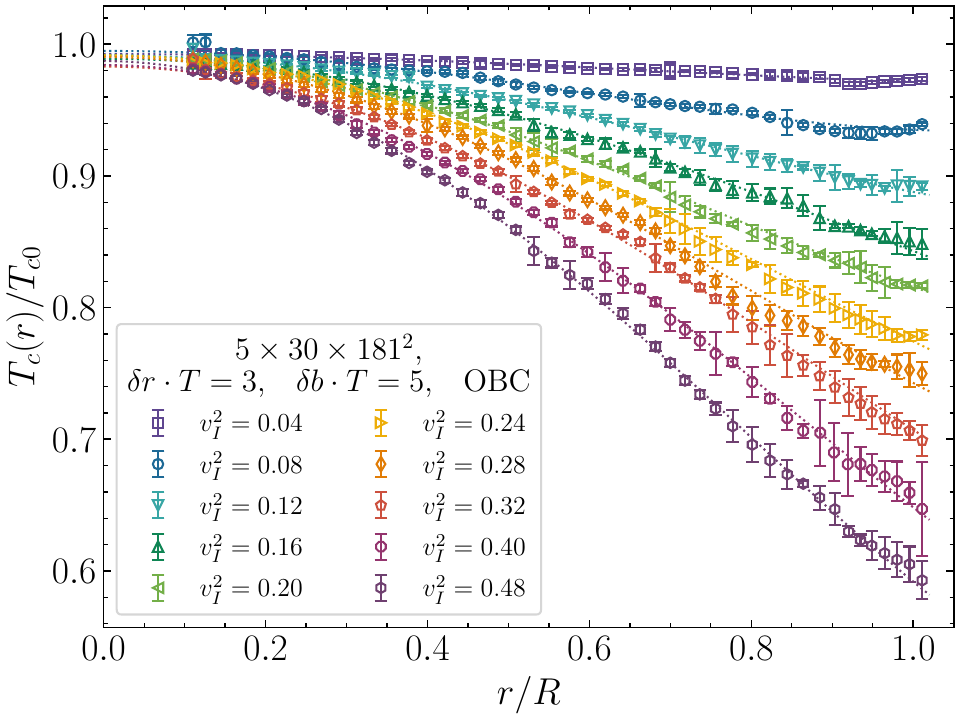}
    \caption{The local critical temperature as a function of radial distance $r$, given in units of the transverse size of the system, for various imaginary velocities $v^2_I$. The results were obtained on the lattice $5\times 30\times 181^2$ with OBC. Dotted lines represent the fit of the data by the equation (\ref{eq_fit_4}).}
    \label{fig_tcloc-compT-nt5}
\end{figure}

Let us now study how the coefficients in the formula (\ref{eq_fit_4}) depend on the velocity of rotation. In Fig.~\ref{fig_tcloc-compT-nt5} we show the local temperature $T_c(r)$ as a function of radius for various imaginary velocities $v_I^2 = 0.04,\,\dots,\,0.48$. These results were obtained for the averaging width $\delta r \cdot T = 3$ and the ignored boundary layer $\delta b \cdot T = 5$ on the lattice $5\times 30\times 181^2$. 

The results shown in Fig.~\ref{fig_tcloc-compT-nt5} are qualitatively similar to those in Fig.~\ref{fig_tcloc-compBC-compW-nt5_1}. However, from Fig.~\ref{fig_tcloc-compT-nt5}, one sees that the behavior of the $T_c(r)$ strongly depends on the rotation velocity $v^2_I$. In particular, the larger the imaginary velocity $v^2_I$, the stronger the dependence on the distance of the local critical temperature $T_c(r)$.

\begin{figure}[t]
    \centering
    \includegraphics[width = 0.60\linewidth]{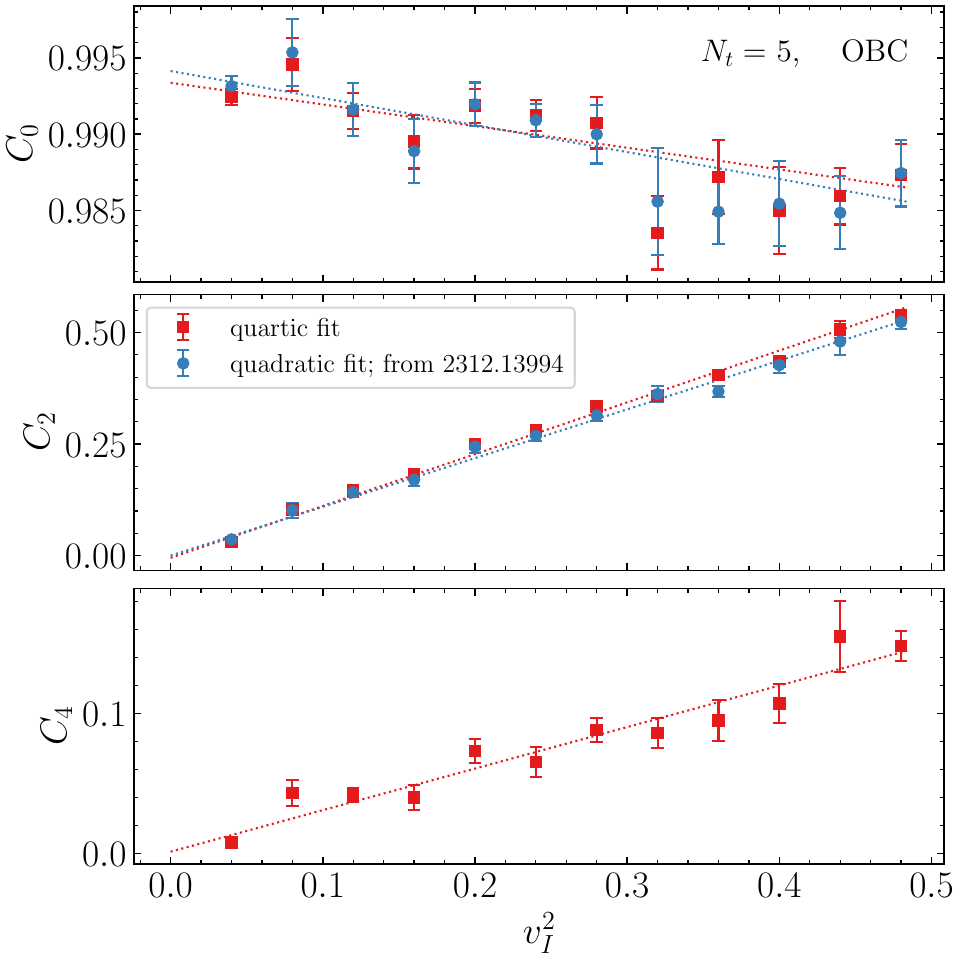}
    \caption{
    The coefficients $C_0, C_2, C_4$ 
    {that encode spatial inhomogeneities of the local critical temperature~\eqref{eq_fit_4} as functions} 
    of $v_I^2$. Results are obtained for open boundary conditions on the lattice $5\times 30\times 181^2$. Dotted lines show linear fits~\eqref{eq_fit_Ci} of the data. The results of the quadratic fit from Ref.~\cite{Braguta:2023iyx} are also presented. The uncertainties include both statistical and systematic contributions.
    }
    \label{fig_coef-compV-nt5}
\end{figure}

In order to determine the coefficients $C_0, C_2, C_4$, the data in Fig.~\ref{fig_tcloc-compT-nt5}  were fitted by formula~(\ref{eq_fit_4}). The quality of these fits is $\chi^2/{\rm n.d.f.} \lesssim 2$. We consider the layer width $\delta r$ as a source of systematic uncertainties and average the values of best-fit coefficients for different $\delta r$ to improve the quality of the data points (see the details in Appendix~\ref{app_Systematic}). The resulting coefficients  $C_0,\, C_2,\, C_4$ are shown in Fig.~\ref{fig_coef-compV-nt5}.

Our data for the $C_0,\, C_2,\, C_4$ can be well described by the linear functions of $v_I^2$:
\begin{subequations} \label{eq_fit_Ci}
\begin{gather}
    C_0(v_I^2) = a_0 - \kappa_0 v_I^2\,, \label{eq_fit_C0} \\
    C_2(v_I^2) = a_2 + \kappa_2 v_I^2\,, \label{eq_fit_C2} \\
    C_4(v_I^2) = a_4 + \kappa_4 v_I^2\,. \label{eq_fit_C4}
\end{gather}
\end{subequations}
from which one can determine the parameters $a_0, a_2, a_4$ and $\kappa_0, \kappa_2, \kappa_4$.

It is clear from formulas (\ref{eq_fit_4})  and  (\ref{eq_fit_Ci}) that the coefficients $a_0, a_2, a_4$, which do not depend on the imaginary angular frequency $\Omega_I$, describe the critical temperature of gluodynamics in the absence of rotation:
\begin{equation}\label{eq_Tc_Omega0} 
    \frac{T_c(r)}{T_{c0}} = a_0 - a_2 \left( \frac{r}{R}\right)^2 + a_4 \left( \frac{r}{R}\right)^4, 
    \qquad [\Omega = 0].
\end{equation}
In non-rotating gluon plasma, the inhomogeneities in the critical temperature may only appear due to the presence of the boundaries. Therefore, for a sufficiently large volume, we expect that the boundary effects in bulk are small, indicating that in the thermodynamic limit, $a_0 \to 1$, $a_2 \to 0$, and $a_4 \to 0$. The deviations of the parameters $a_0, a_2, a_4$ from these values would signal the presence of finite volume effects. In our calculation, we found that the parameters $a_2$ and $a_4$ are indeed zero within a small numerical uncertainty while the deviation of the leading coefficient $a_0$ from unity is smaller than $1\%$ (cf. Fig.~\ref{fig_coef-compV-nt5}). These properties undoubtedly demonstrate that finite volume effects are small and under control in our study.

\begin{figure}[t]
    \centering
    \includegraphics[width = 0.60\linewidth]{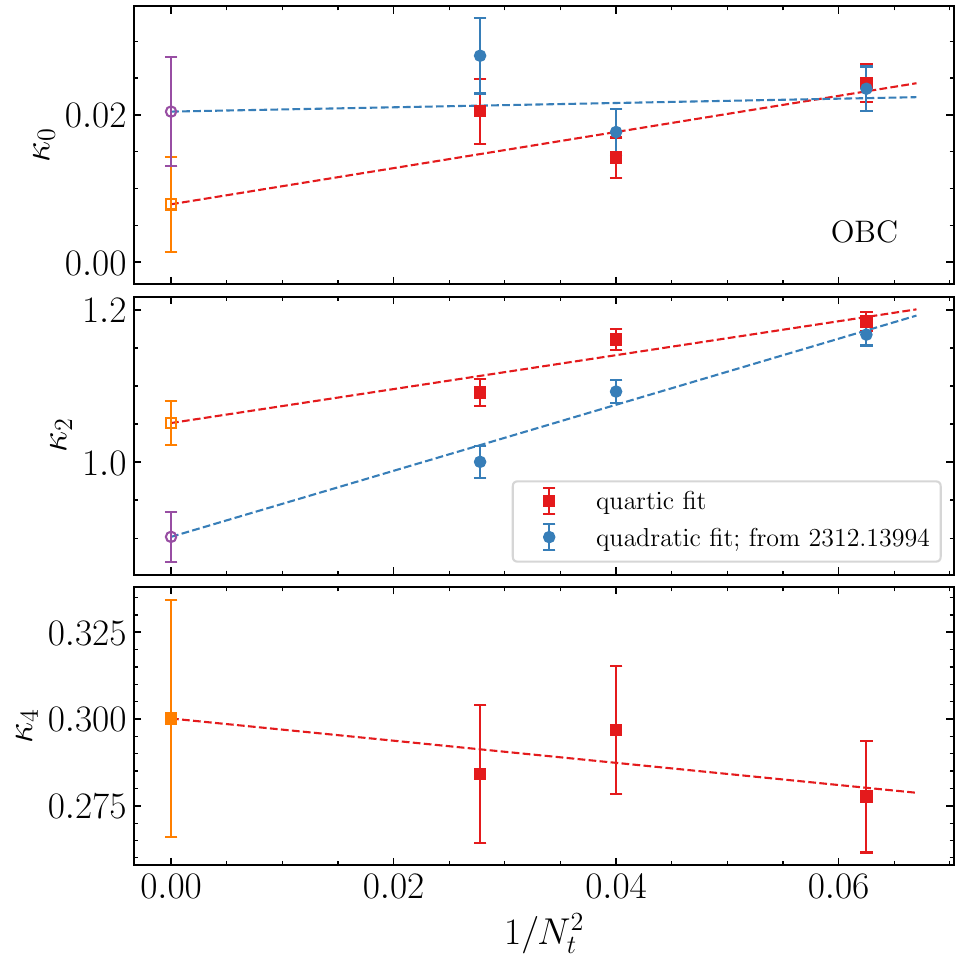}
    \caption{
    The continuum limit extrapolation for the fit coefficients $\kappa_i$ in Eq.~\eqref{eq_fit_Ci}.
    In addition, we show the results from our first study~\cite{Braguta:2023iyx} (the blue markers).
    }
    \label{fig_k-continuum}
\end{figure}

Given a moderate dependence of the critical temperature on the boundary conditions, it is important to determine the properties of the vortical hot gluonic matter at lattices that approach the continuum limit. To find the continuum limit for the coefficients $\kappa_0$, $\kappa_2$, $\kappa_4$ in addition to the lattice $5\times 30\times 181^2$ we repeated our calculations on the lattices $4\times 24\times 145^2$, $6\times 36\times 217^2$ that have different size in lattice units as well as different lattice spacings. The results found in this way are presented in Fig.~\ref{fig_k-continuum}. 

The continuum limit for the $\kappa_0$ appears to be zero within the small numerical uncertainties. Taking into account the formula (\ref{eq_fit_C0}), this property implies that the coefficient $C_0$ is unity up to the finite volume effects. Noticing that the coefficient $C_0$ determines the ratio $T_c(r=0)/T_{c0}$ for rotating gluodynamics we conclude that {\it the critical deconfining temperature on the rotating axis $T_c(r=0)$ for rotating gluon matter coincides with the critical temperature of the same system without rotation.} This result should be confronted with perturbative estimations of Ref.~\cite{Chen:2022smf} where such dependence has been found.

Now let us consider the vortical coefficients $\kappa_2$ and $\kappa_4$ {that appear in Eqs.~\eqref{eq_fit_C2} and \eqref{eq_fit_C4}, respectively.}
The continuum limit for these coefficients, obtained through the quartic fit of the data, gives us the following values: 
\begin{equation}\label{eq_res_kap24}
    \kappa_2 = {1.051(29)}\,,\qquad \kappa_4 = {0.300(34)}\,.
\end{equation}
It is worth comparing this result to the value 
\begin{equation}\label{eq_res_kap2}
    \kappa_2 = {0.902(33)}\,,
\end{equation}
which has been obtained in Ref.~\cite{Braguta:2023iyx} through the quadratic ($C_4=0$) fit of the data in the bulk with $a_2=0$. We also found that the rotation-dependent correction $\kappa_0$ to the on-axis critical temperature $C_0$, given by Eq.~\eqref{eq_fit_C0}, has a vanishingly small value, $\kappa_0 = 0.008(6)$.

To summarize the results of this section, the critical temperature for rotating gluodynamics can be well described by the following formula\footnote{Previous lattice studies~\cite{Braguta:2020biu, Braguta:2021jgn, Braguta:2021ucr, Braguta:2022str} have found a dependence of the critical temperature $T_c$ on $\Omega_I$ without accounting for a dependence of the phase structure on the distance from the center of rotation. These findings should be interpreted as bulk-averaged critical temperatures.}:
\begin{equation}\label{eq_fit_4_result} 
    \frac{T_c(r \Omega_I )}{T_{c0}} = 1 - \kappa_2 (\Omega_I r)^2 + \kappa_4 (\Omega_I r)^2 \left( \frac{r}{R}\right)^2\,.
\end{equation}
Our data suggests that the coefficient $\kappa_2$ is universal in bulk, i.e., it does not depend on the lattice parameters such as the transverse lattice size and the type of boundary conditions. On the contrary, the quartic coefficient $\kappa_4$ depends on these parameters of the system.

Equation~(\ref{eq_fit_4_result}) can be rewritten as follows:
\begin{equation}\label{eq_fit_4_result_1} 
    \frac{T_c(r,\Omega_I )}{T_{c0}} = 1 - (\Omega_I r)^2 \left ( \kappa_2  - \kappa_4  \left( \frac{r}{R}\right)^2 \right ) \,.
\end{equation}
In this form, the quartic $\kappa_4$-term resembles a finite volume correction to the $\kappa_2$ coefficient. In addition to the observed $\kappa_4$-term, the $r^4$--dependence of the critical temperature can be parameterized by the term $\sim (\Omega_I r)^4$. In our simulations, the dominant contribution results from the $\kappa_4$ term, and, with our numerical accuracy, we are not able to distinguish it from the suspected $\sim (\Omega_I r)^4$ term. However, one can expect that there still exists a contribution of the form $\sim (\Omega_I r)^4$ to the critical temperature, which will be important for sufficiently large angular velocity. We will discuss this $(\Omega_I r)^4$ contribution in Section \ref{sec_Local} within the approximation of local thermalization.

\subsection{The phase diagram for real and imaginary rotation}

We have already discussed that for imaginary rotation, the mixed inhomogeneous phase can be observed for the on-axis temperatures below the critical temperature of non-rotating gluodynamics: $T=T_{c0}-\Delta T$ with $\Delta T>0$. In this case, the confinement/deconfinement phases are localized to the central/periphery regions of the studied volume. The critical distance $r$, where the transition from the confinement to the deconfinement takes place, can be found from the equation: $T_{c0}-\Delta T = T_c(r,\Omega_I )$, where the $T_c(r,\Omega_I )$ is given by (\ref{eq_fit_4_result}) and $\Omega_I$ and $\Delta T$ serve as the parameters of this equation.  

To carry out the analytical continuation of the phase diagram from imaginary to real rotation, we make the following substitution:
\begin{align}
	\Omega^2_I \to -\Omega^2\,.
 \label{eq_Omega_an}
\end{align}

\begin{figure}[t]
    \centering
    \includegraphics[width = 0.60\linewidth]{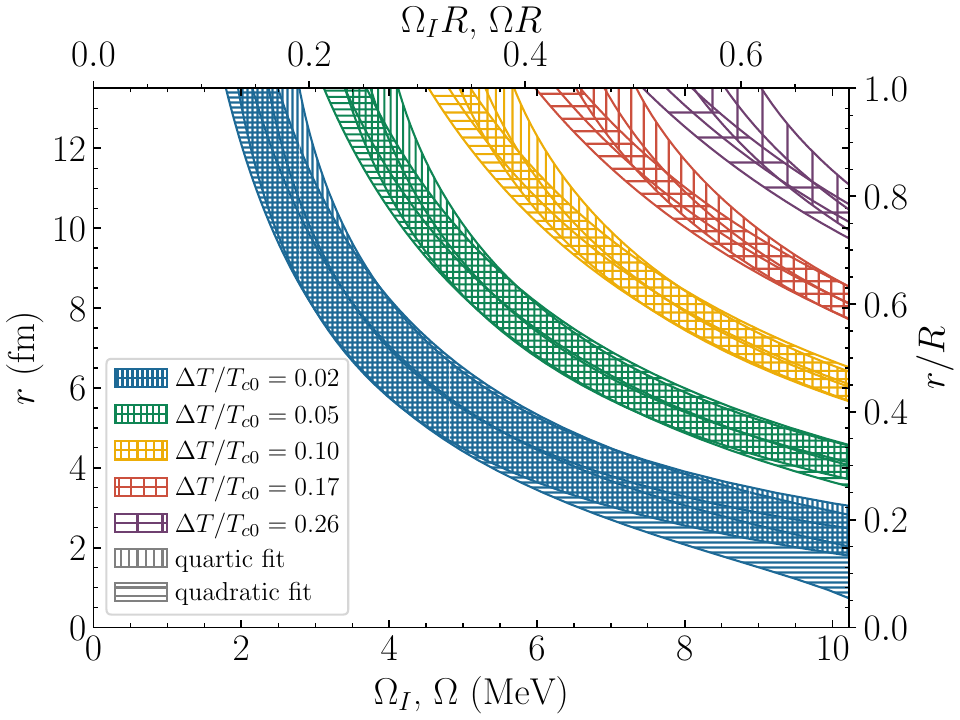}
    \caption{The critical radius $r$ of the spatial boundary between the confining and deconfining phases for, simultaneously, the imaginary frequencies $\Omega_I$ (on-axis temperature $T = T_{c0} - \Delta T$) and the real frequencies $\Omega$ (on-axis temperature $T = T_{c0} + \Delta T$) for various temperature offsets $\Delta T >0$. The shading corresponds to a degree of uncertainty around the central curve. The size of the system is $R = 13.5$~fm, which is typical for our lattice simulations. The regions with the horizontal hatching represent the diagram from Ref.~\cite{Braguta:2023iyx}, obtained with the help of a quadratic fit [i.e., with $\kappa_4 \equiv 0$ in Eq.~\eqref{eq_fit_4_result}].
    }
    \label{fig_diagram_OmegaI}
\end{figure}

Following the analytical continuation~\eqref{eq_Omega_an} applied to Eq.~\eqref{eq_fit_4_result}, one finds that for the rotation with a real angular velocity, the deconfinement phase is localized in the central region, while the confinement phase is in the periphery of the system. As we mentioned in Ref.~\cite{Braguta:2023iyx} and also will discuss below, this is an exotic phenomenon that is totally unexpected from generic considerations based on the Tolman-Ehrenfest law. The critical distance from the rotation axis, where the spatial deconfinement/confinement transition appears, is the solution of the following equation: $T_{c0}+\Delta T = T_c(r,\Omega)$. This procedure gives us the phase diagram, which is shown in Fig.~\ref{fig_diagram_OmegaI}.\footnote{In constructing the diagram in Fig.~\ref{fig_diagram_OmegaI}, we include a minor deviation of $a_0$ from unity due to finite volume effects to uncertainty.}
Note that Eq.~\eqref{eq_fit_4_result} is a quadratic function in $\Omega_I$. Therefore, for a given temperature offset $\Delta T$, the diagram for the real angular velocity at the temperature $T_{c0} + \Delta T$ has the same shape as for the imaginary rotation at the symmetric values of temperature $T_{c0} - \Delta T$.

In Fig.~\ref{fig_diagram_OmegaI}, we also plot the diagram using the results of quadratic fit (i.e. formula~\eqref{eq_fit_4_result} without $\kappa_4$-term) from Ref.~\cite{Braguta:2023iyx}.
The minor difference between the quadratic and quartic results appears for large velocities of rotation, mostly near the boundaries $r/R \gtrsim 0.8$. Nevertheless, these results are in agreement with each other within the statistical uncertainties.

\section{Decomposition of rotating action} \label{sec_Decomposition}

\subsection{Mechanical vs. chromomagnetic vortical effects.}

It is worth asking what is the origin of the peculiar behaviour of gluon plasma subjected to rotation? More concretely, which terms in the action are responsible for the statistical properties of gluonic gas? To answer this question below, we decompose the action into mechanical and magnetic terms, similar to Eq.~\eqref{eq_L_G_decomposition}, and analyze their effects associated with rotation separately.

Similarly to Lagrangian~\eqref{eq_L_G_decomposition}, the action of rotating gluodynamics (\ref{eq_S_E_continuum_c}) is a quadratic function in {the imaginary angular frequency} $\Omega_I$:
\begin{equation} \label{eq_action_decomposition}
    S(\Omega_I)= S_0 + S_1 \Omega_I + S_2 \Omega_I^2\,, 
\end{equation}
where $S_0$ is a well-known action in gluodynamics without rotation. The contributions $S_1$ and $S_2$ in the cylindrical coordinates can be written as
\begin{align}
S_1 \equiv S_{\rm mech} & = -\frac {1} {g_{YM}^2} \int\! d^4x\,r \left[ 
 F^a_{\hat \varphi r} F^a_{r \tau} + F^a_{\hat \varphi z} F^a_{z \tau} 
\right], 
\label{eq_S1_mech}\\
S_2 \equiv S_{\rm magn} &= \frac {1} {2g_{YM}^2} \int\! d^4x\,r^2 \left[ 
(F^a_{\hat \varphi z})^2 + (F^a_{r \hat \varphi })^2
\right].
\label{eq_S2_magn}
\end{align}
The terms $S_1$ and $S_2$ have a clear physical meaning~\cite{Braguta:2023kwl}. The first term, $S_1$, is related to the total angular momentum of gluons and, according to the discussion in Section~\ref{sec_Couplings}, represents a linear ``mechanical'' coupling with the angular velocity. The second term, $S_2$, is given by the squares of two components of chromomagnetic field $(B_r^a)^2$ and $(B_z^a)^2$. This term represents a quadratic ``chromomagnetic'' coupling with the angular velocity, which stresses the highlighted role of the chromomagnetic condensate in the rotational motion of the gluon plasma.

In papers~\cite{Braguta:2023yjn, Braguta:2023kwl, Braguta:2023qex, Braguta:2023tqz}, the moment of inertia of gluon plasma was studied. It was found that the mechanical coupling $S_1$ and the chromomagnetic coupling $S_2$ have opposite influences on the moment of inertia. In particular, the term that couples to the angular velocity linearly, $S_1$, gives a positive contribution to the moment of inertia. On the contrary, due to the specific behavior of magnetic gluon condensate, the contribution of the term which couples to the angular frequency quadratically, $S_2$, is negative up to the temperatures $\sim 2 T_c$. It was also suggested 
that these operators pull the critical temperature in opposite directions: for imaginary rotation, the linear mechanical coupling to rotation $S_1$ leads to an increase of the critical temperature in agreement with our intuition, while the quadratic chromomagnetic coupling $S_2$ tends to decrease the critical temperature.

In this section, we are going to study the roles of the $S_1$ and $S_2$ operators in the mixed inhomogeneous phase of rotating gluodynamics. To do this we introduce the factors $\lambda_1$ and $\lambda_2$ to the action (\ref{eq_action_decomposition}):  
\begin{equation}\label{eq_S_E_labmdas}
    S_G = S_0 + \lambda_1 S_1 \Omega_I + \lambda_2 S_2 \Omega_I^2\,.
\end{equation}
These factors, which take the values $\lambda_{1} = 0,1$ and $\lambda_{2} = 0, \pm 1$, allow us to switch on/off the $S_1$ and $S_2$ operators in the simulation. In particular, the following regimes are possible:
\begin{itemize}
    \item[Im1)] 
    $\lambda_1 = 1,\ \lambda_2 = 0$ {(the linear coupling only)}: The imaginary rotation acts only through the linear term $S_1$;
    \item[Im2)] 
    $\lambda_1 = 0,\ \lambda_2 = 1$ {(the quadratic coupling only)}:  The imaginary rotation acts only through the quadratic term $S_2$;
    \item[Im12)] 
    $\lambda_1 = 1,\ \lambda_2 = 1$ {(the full action)}:
    This case, {which corresponds to the imaginary rotation of the whole action}, does not require additional study since it has been considered in Section~\ref{sec_Coexistence}. 
\end{itemize}
Note that in the case of Im2, there is, actually, no sign problem. Therefore, for this choice of the factors $\lambda_i$, we can conduct lattice simulation with real angular velocity. In order to study this case, we introduce one more regime
\begin{itemize}
    \item[Re2)] $\lambda_1 = 0$, $\lambda_2 = - 1$ (real rotation for the quadratic coupling): The rotation acts only through the quadratic term $S_2$, but due to the negative value of the coupling, the angular velocity takes a real value: $\Omega^2 = - \lambda_2 \Omega_I^2 >0$.
\end{itemize}

We stress that the separation of the action into three different parts~\eqref{eq_S_E_labmdas} does not violate the gauge invariance of the action since all three contributions, $S_0$, $S_1$ and $S_2$ are invariant under SU(3) gauge transformations.

\subsection{Inhomogeneous phases}

\begin{figure*}
    \includegraphics[width = 1.0\linewidth]{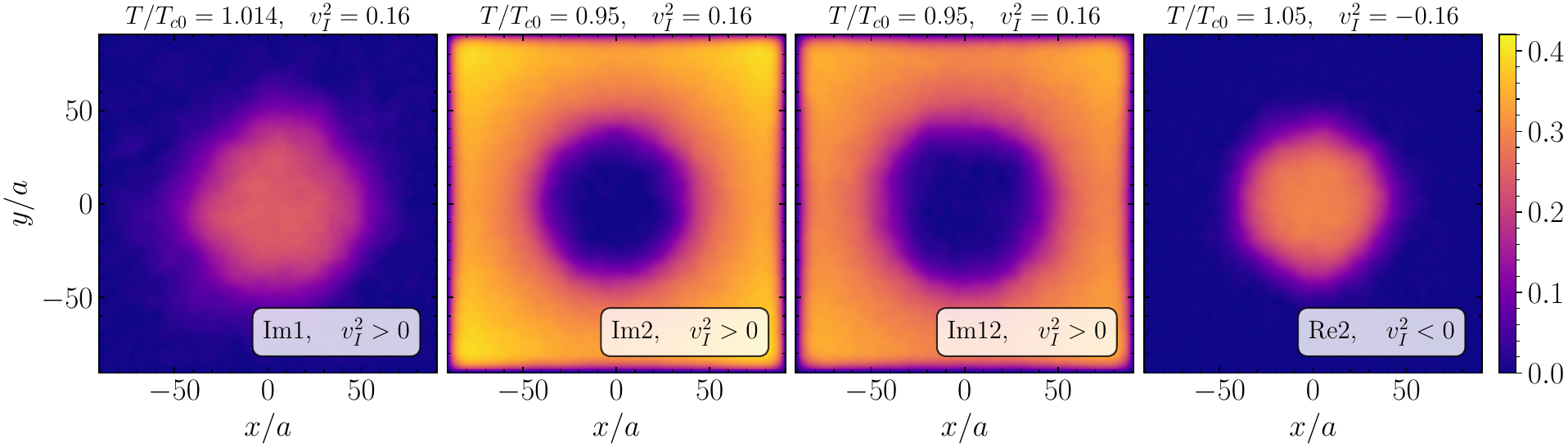}
    \caption{The distribution of the local Polyakov loop in $x,y$-plane for the lattice size $5\times 30\times 181^2$ with OBC at the fixed absolute value of rotational velocity $|v^2| = |\Omega R|^2 = 0.16$ and various regimes corresponding, from the left plot to the right plot to the linear imaginary coupling (Im1), the quadratic imaginary coupling (Im2), the full action in rotation (Im12), and the real angular velocity acting only the quadratic-coupling term (Re2), respectively. On each panel, the temperatures are chosen in such a way that the mixed phase occurs. Movies on the phase evolution with increasing temperature for different regimes are available as ancillary files (see Supplementary material).
    }
    \label{fig_imshowPLxy-v4-lambdas}
\end{figure*}

The distributions of the local Polyakov loop for different regimes Im1, Im2, Im12, and Re2 are shown in Fig.~\ref{fig_imshowPLxy-v4-lambdas}. The simulations were carried out at the fixed value of the velocity $v_I^2 = 0.16$ for the Im1, Im2, Im12 and $v_I^2 =-0.16$ for the Re2 on the lattice $5\times 30 \times 181^2$ with OBC. We have adjusted the temperatures on the central axis (indicated in the plots) to observe the mixed-phase structures. The movies with the phase evolution in different regimes for both OBC and PBC are attached as ancillary files (see Supplementary material).

\vskip 1mm
\paragraph*{\bf Linear coupling at imaginary rotation.}
In the first regime, Im1, the mixed-phase structure can be observed for temperatures $T \gtrsim T_{c0}$, which are slightly above the critical temperature of non-rotating gluodynamics. This mixed phase arises only in a narrow temperature range because the effects of linear coupling are relatively weak (in the left panel of Fig.~\ref{fig_imshowPLxy-v4-lambdas}, we show the local Polyakov loop distribution for the temperature $T/T_{c0} = 1.014$, for which the mixed phase is clearly visible). However, in this case, the deconfinement phase is localized in the center region, whereas the confinement phase arises at the periphery of the system. As one increases the temperature at the rotation axis, the deconfinement phase captures larger and larger volumes. Finally, at some temperature, the whole system transits to the deconfinement phase; for the temperatures, $T \lesssim T_{c0}$, the mixed-phase structure does not appear. Thus, the mixed-phase structure of the Im1-regime has properties opposite to that observed in Section~\ref{sec_Coexistence}, where the imaginary rotation with the full action (the case Im12) has been studied.

\paragraph*{\bf Quadratic coupling at imaginary rotation.}
A different picture is observed for the Im2-regime, {where the rotation couples only to the chromomagnetic component of the gluon field}. In this case, the mixed-phase structure has properties qualitatively similar to that of rotating gluodynamics (Im12-regime  studied in Section~\ref{sec_Coexistence}): the confinement in the central region and the deconfinement at the periphery. However, quantitatively, the Im2-regime differs a bit from the regime Im12 (see two central panels in Fig.~\ref{fig_imshowPLxy-v4-lambdas}, where the phase structures are shown for these regimes at the same value of the (on-axis) temperature $T/T_{c0} = 0.95$).

\paragraph*{\bf Quadratic coupling at real rotation.}
The Re2-regime is designed to realize real rotation for the $S_2$ term. For a fixed $| v_I |$, the gluon matter has properties opposite to those of the Im2-regime: the deconfinement arises in the central region, and the confinement is located at the periphery. In addition, in the Re2-regime, the mixed phase appears for the temperatures $T>T_{c0}$, which differ from the mixed-phase domain in the Im2-regime, $T<T_{c0}$. In the right panel of Fig.~\ref{fig_imshowPLxy-v4-lambdas} the distribution of the local Polyakov loop is shown for $T/T_{c0} = 1.05$. 

\begin{figure}[t]
    \centering
    \includegraphics[width = 0.60\linewidth]{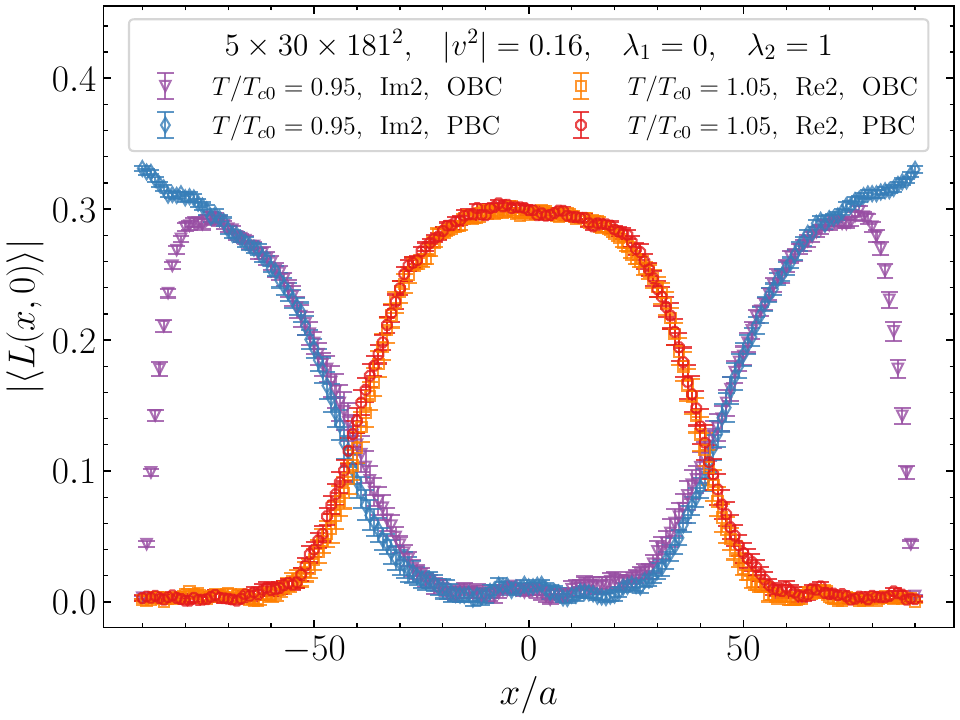}
    \caption{The effect of the quadratic-only coupling $S_2$ of the vorticity to the gluon action:
    The distribution of the local Polyakov loop as a function of $x$ coordinate for lattice size $5\times 30\times 181^2$, open (OBC) and periodic (PBC) boundary conditions for the Im2 and Re2 regimes. Data are shown, respectively, for $|v^2| =  0.16$ at $T/T_{c0} = 0.95$ with imaginary angular velocity (Im2) and $T/T_{c0} = 1.05$ with real angular velocity (Re2).
    }
    \label{fig_compReIm-PLx-lambdas}
\end{figure}

To demonstrate the opposite behavior of the regimes Im2 and Re2, in  Fig.~\ref{fig_compReIm-PLx-lambdas}, we show the local Polyakov loop as a function of $x$ coordinate for both regimes and OBC/PBC. The results were obtained for the positive (negative) shift of the on-axis temperature $\Delta T = 0.05\, T_{c0}$ and $|v_I^2|=0.16$, i.e. $T/T_{c0} = 0.95,\, v_I^2 = 0.16$ and $T/T_{c0} = 1.05,\, v_I^2 =-0.16$.
One sees that the confinement phase translates into the deconfinement phase after analytic continuation and vice versa. The boundary between the phases remains at the same position. Notice also that the spatial boundary conditions affect the results only in the regions close to the boundary within a finite-temperature correlation length.

\begin{figure}[t]
    \centering
    \includegraphics[width = 0.60\linewidth]
    {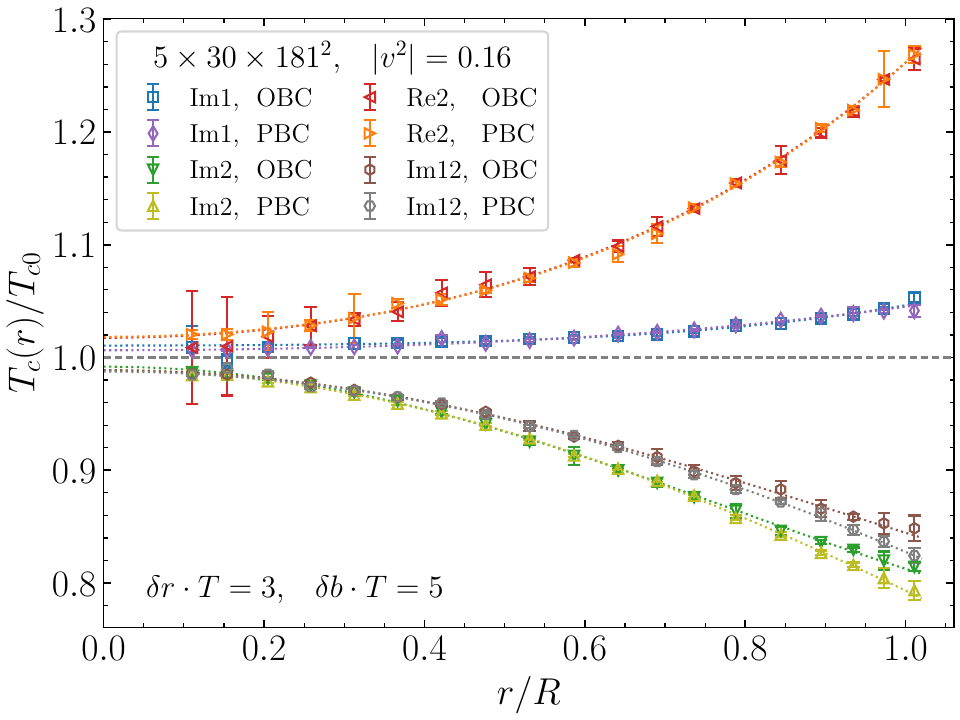}
    \caption{The local critical temperature as a function of radial distance for different regimes of rotation.
    The results from the lattice $5\times 30\times 181^2$ with open (OBC) and periodic (PBC) boundary conditions at fixed rotation velocity $|v^2| = |\Omega R|^2 = 0.16$.
    Dotted lines represent the fit of the data by the function~\eqref{eq_fit_4}.}
    \label{fig_tcloc-compLambda}
\end{figure}

\paragraph*{\bf Phase boundaries.}
Using the methods developed in the previous section, we compute the local critical temperatures for all regimes considered in this section. The results of this calculation are shown in Fig.~\ref{fig_tcloc-compLambda}. It is seen from this plot that the rotation in regime Im1 increases the critical temperature of gluodynamics, whereas in regime Im2, it decreases due to the rotation. This result confirms our suggestion proposed in paper~\cite{Braguta:2023yjn} that the term with total gluon angular momentum $S_1$ increases, while the chromomagnetic field contribution $S_2$ decreases the critical temperature. It is also seen from Fig.~\ref{fig_tcloc-compLambda} that the radial dependence of the $T_c(r)$ for the Im1-regime is very weak.
For this reason, the local critical temperatures of the regimes Im2 and Im12, which differ by the accounting of $S_1$ term, are close to each other.  The next points to be discussed are the Im2- and Re2- local critical temperatures. Our results for the critical temperature indicate that within the uncertainty, the results for regimes Im2 and Re2 are connected through the analytical continuation procedure $\Omega^2 \leftrightarrow -\Omega_I^2$ in Eq.~\eqref{eq_fit_4_result}.\footnote{This agreement is excellent up to the large radial distances $r/R \sim 0.8$; the minor discrepancy appears only in the vicinity of the boundary. It may indicate that the next-order terms may be needed to reconstruct the boundary effects fully after analytic continuation.}

Briefly summarizing this section, we conclude that the standard linear coupling of the vorticity to the angular momentum of the gluons, represented by the action $S_1$, Eq.~\eqref{eq_S1_mech}, leads to an increase of the critical temperature of the deconfinement phase transition as the imaginary angular frequency increases. This effect is anticipated from earlier analytical studies~\cite{Chernodub:2020qah} and, at the same time, it is opposite to the first-principle results of lattice simulations of the full gluonic action~\cite{Braguta:2020biu, Braguta:2021jgn}. The reason why the imaginary rotation leads to the decrease of the critical temperature in fully-rotated gluodynamics appears to be related to the dominance of the quadratic coupling of the vorticity to the chromomagnetic component of the gluon fields, which is characterized by the action $S_2$, Eq.~\eqref{eq_S2_magn}. This result also agrees with our earlier finding that the mechanical properties of the rotating gluon plasma are predominantly influenced by the chromomagnetic condensate~\cite{Braguta:2023yjn, Braguta:2023tqz}.

\section{Local thermalization of rotating gluodynamics}\label{sec_Local}

From a simple kinematic point of view, the influence of the rotation on the gluonic fields has two major effects: Yang-Mills action~\eqref{eq_S_E_continuum_c} becomes both inhomogeneous and anisotropic. 
In details, 
\begin{itemize}
\item[(i)] Inhomogeneity: the gluonic action develops an explicit dependence of the coefficients in front of various gluonic operators on the distance from the rotation axis $r$;
\item[(ii)] Anisotropy: The rotation affects chromoelectric and chromomagnetic components of gluonic fields differently.
\end{itemize}
In the previous sections, we studied both effects simultaneously. In this section, we will concentrate mostly on the impact of anisotropy on the gluonic action in the rotating reference frame.

\subsection{Local approximation for inhomogeneous action}
The results presented in the previous sections were obtained on the lattices with quite large extensions in the transverse directions, which complicates the simulation considerably.  The study of rotating gluodynamics on large lattices is necessary since we put the axis of rotation to the center of the lattice and thus effectively half the lattice sizes in the $x$- and  $y$-directions. Furthermore, the phenomena studied in this paper cannot be observed on small lattices because of the rather wide confinement/deconfinement transition region and boundary effects.  Another property which complicates the study of rotating gluodynamics is 
an explicit dependence of its action (\ref{eq_S_E_continuum}), (\ref{eq_S_E_continuum_c}) on the coordinate $r$, which complicates theoretical study of this system. 

In order to simplify the situation from both numerical and theoretical perspectives, we introduce the following approximation. Let us consider the system at distances $r \sim r_0 \gg \zeta$ from the rotation action, where $\zeta$ is the correlation length of gluodynamics. It is clear that in this approximation, the coefficients of the gluon fields operators in action (\ref{eq_S_E_continuum_c}), which depend on the product $\Omega r$, remain approximately the same if one varies the distance by few $\zeta$ in the vicinity of $r_0$. Therefore, it is reasonable to assume that in this region, thermodynamic parameters of the system correspond to the action  (\ref{eq_S_E_continuum_c}) with the coefficients fixed to their values at a distance $r=r_0$. One can expect that this idea works for small distances $r_0 \lesssim \zeta$ as well. This property can be seen as follows: if one considers slow rotation $\Omega_I \zeta \ll 1$, which is always valid in our simulations, the coefficients preceding various terms in action~(\ref{eq_S_E_continuum_c}) do not change considerably. One could expect a prevailing contribution coming from the linear $\Omega_I r$ term over the terms that are quadratic in $\Omega_I r$. However, according to the results of the previous section, this linear term does not play an important role for the spatial layers located at small distances from the rotation axis. 

In this section, we are going to study rotating gluodynamics through the simulation of the action (\ref{eq_S_E_continuum_c}) with the radius $r$ fixed at some value $r_0$. In this case, the action does not depend on the coordinate explicitly, and the system becomes homogeneous.  Below, we will call this approximation as {\it local thermalization}. Actually, one can fix the coefficients at any point on the circle $r=r_0$. In our simulation, we choose the point $x=r_0,~ y=0$. The local velocity at this point is $u_I=\Omega_I r_0$, and the homogeneous local action to be simulated can be written as
\begin{multline}\label{eq_S_E_continuum_local}
	S_{G} = \frac{1}{2 g_{YM}^{2}} \int\! d^{4}x \ \Big[
    F^a_{x \tau} F^a_{x \tau} + F^a_{y \tau} F^a_{y \tau} + F^a_{z \tau} F^a_{z \tau} 
    + F^a_{x z} F^a_{x z} + {}\\
    + \left(1 + u_I^2\right) F^a_{y z} F^a_{y z} 
    + \left(1 + u_I^2\right) F^a_{x y} F^a_{x y} 
    - 2 u_I \left(F^a_{y x} F^a_{x \tau} + F^a_{y z} F^a_{z \tau}\right)
    \Big]\, .
\end{multline}
Notice that in this approximation, the observables depend on the $u_I$, and they are equal to the ones in non-rotating gluodynamics at $u_I=0$.

Lattice simulation with the action (\ref{eq_S_E_continuum_local}) can be carried out using standard lattice methods with periodic boundary conditions in all directions. 
We discretize the local action~\eqref{eq_S_E_continuum_local} in the same way as the action of full rotating system~\eqref{eq_S_E_continuum} (see Section~\ref{sec_Formulation_setup}).
We determine the critical lattice coupling from the peak of the Polyakov loop susceptibility. Then, using the $\beta$ function of non-rotating gluodynamics, we calculate the critical temperature $T_c$, which can be interpreted as a critical temperature on the axis of rotation.

\subsection{The results obtained with the local action}\label{sec_Local_results}

\begin{figure}
    \centering
    \includegraphics[width = 0.60\linewidth]{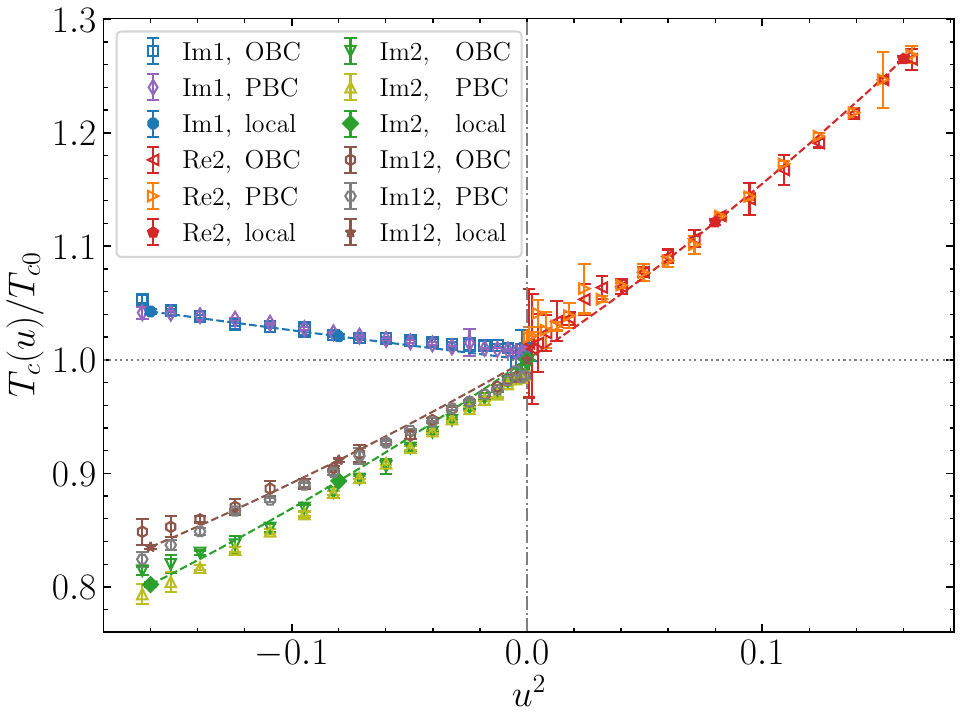}
    \caption{
    The critical (on-axis) temperature as the function of the local velocity $u^2 = -u_I^2$ for various regimes: Im1, Im2, Im12, Re2. The data for open (OBC) and periodic (PBC) boundary conditions are calculated in Section~\ref{sec_Decomposition} on the lattice $5\times 30\times 181^2$. The data obtained within the local thermalization approximation are calculated on the lattice $5\times 20^3$. The dashed curves represent the fitting of the data by the formula~(\ref{eq_fit_u_4}). 
    }
    \label{fig_tcloc-u-compLambda}
\end{figure}

First, we carry out the study of the local thermalization approximation on the lattice $5\times 20^3$. Note that the action~\eqref{eq_S_E_continuum_local} supports the decomposition using Eq.~\eqref{eq_S_E_labmdas} and the Im1-, Im2-, Im12-, Re2- regimes  proposed in the previous section can be realized. In Fig.~\ref{fig_tcloc-u-compLambda}, we show the results for the (on-axis) critical temperature as a function of the $u^2 = -u_I^2$ calculated for different regimes. For comparison, the results from Fig.~\ref{fig_tcloc-compLambda}, which were computed on the lattice $5\times30\times181^2$ with the action (\ref{eq_S_E_continuum}),  are also shown.
It is seen from Fig.~\ref{fig_tcloc-u-compLambda} that the results for the actions~(\ref{eq_S_E_labmdas}) and (\ref{eq_S_E_continuum_local}) are in good agreement with each other in the shown region $|u^2|\lesssim 0.2$. This result confirms that for sufficiently small velocity $u^2$, the local thermalization approximation works within the uncertainty of the calculation for all the regimes. 

In the studied region of rotational velocity $|u^2| \lesssim 0.5$ the data are well described by the polynomial:
\begin{equation} \label{eq_fit_u_4} 
    \frac{T_c(u)}{T_{c0}} =
    1 + k_2 u^2 + k_4 u^4\,.
\end{equation}
In Eq.~\eqref{eq_fit_u_4}, we denote the vortical curvature $\kappa$ by $k$ to distinguish the methods. Here, a few comments are in order. First, at $u_I^2=0$, the system under study coincides with non-rotating gluodynamics; therefore, ${T_c(u=0)}/{T_{c0}}=1$. Next, although in the equations (\ref{eq_fit_4_result}) and (\ref{eq_fit_u_4}) the $\kappa_2$ and $k_2$ are similar, the terms $\kappa_4$ and $k_4$ have different physical meanings. The former results from boundary contribution to the $\kappa_2$ coefficient,  while the latter describes the influence of rotation on the critical temperature. We would like to reiterate that the data from Section \ref{sec_Coexistence} does not allow us to find the term analogous to the $k_4$. However, in the local thermalization approximation, which is free from boundary effects, the $k_4$ term can be studied.

The simulation of rotating gluodynamics in the Re2-regime allows one to carry out a lattice study with real angular velocity. Notice that this is not the rotation of the full action~(\ref{eq_S_E_continuum_local}), since in this regime, we disregard the linear in $\Omega$ term. Therefore, it is rather difficult to transfer the conclusions from regime Im2 to physical regime Im12, especially in the region of large velocities $u^2 \sim 1$. However, the Re2-regime is straightforwardly related to the Im2-regime through the procedure of analytical continuation. Thus, one can inspect this procedure for different analytic functions. 

\begin{figure}
    \centering
    \includegraphics[width = 0.60\linewidth]{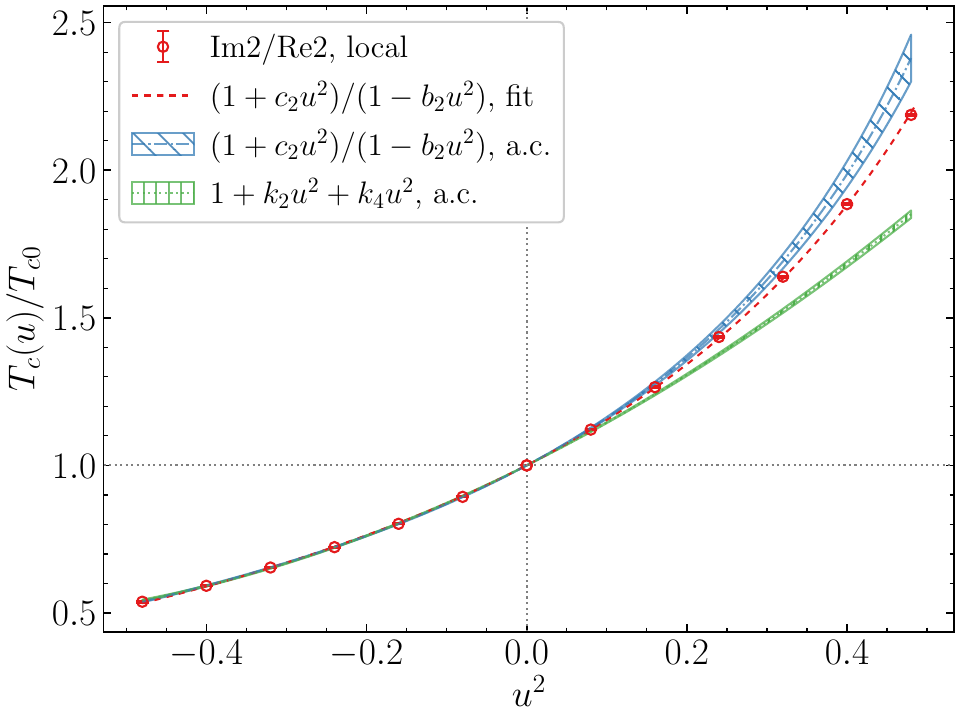}
    \caption{
    The critical temperature for the Im2- and Re2-regimes as a function of $u^2$ (the red open circles). The region $u^2<0$ corresponds to the rotation with imaginary angular velocity realized in the Im2-regime. The region $u^2>0$ represents the real rotation in the Re2-regime. The calculation was conducted on the lattice $5\times 20^3$.
    The hatched regions represent an analytic continuation of the results from region $u^2 < 0$ to region $u^2 > 0$, based on the functions~\eqref{eq_fit_u_4} and~\eqref{eq_fit_u_4_rational}.
    The red dashed line shows the best fit of all the data by the rational function~\eqref{eq_fit_u_4_rational}.
    }
    \label{fig_tcu-Im2Re2-check-continuation}
    \vspace{1.0em}
\end{figure}

In Fig.~\ref{fig_tcu-Im2Re2-check-continuation}, we show the critical temperatures as a function of $u^2$ for these regimes. The region $u^2<0$ corresponds to the rotation with imaginary angular velocity realized in the Im2-regime, while the region $u^2>0$ represents a real rotation in the Re2-regime. 

To fit the data shown in Fig.~\ref{fig_tcu-Im2Re2-check-continuation} in the region $u^2<0$, we tried to implement the formula~(\ref{eq_fit_u_4}). Then, we continue the result, Eq.~\eqref{eq_fit_u_4}, to the domain of real velocity $u^2 > 0$ (the green hatched region in Fig.~\ref{fig_tcu-Im2Re2-check-continuation}) and compare it with lattice data for regime Re2. It turns out that the analytic continuation can describe the data only up to $u^2\lesssim 0.2$. The addition of the next order $k_6 u^6$ term does not improve the situation considerably.

In our analysis, we found that the rational function with a pole structure,
\begin{equation}\label{eq_fit_u_4_rational}
   \frac{T_c(u)}{T_{c0}} = \frac{1 + c_2 u^2}{1 - b_2 u^2}\,,
\end{equation}
can predict the data for $u^2>0$ from the results for $u^2 < 0$ reasonably well (see blue hatched region in Fig.~\ref{fig_tcu-Im2Re2-check-continuation}). We also note that this rational function describes all data from regimes Im2 and Re2 simultaneously (see red dashed line in Fig.~\ref{fig_tcu-Im2Re2-check-continuation}), whereas the polynomial function~\eqref{eq_fit_u_4} is applicable only for each subset of data $u^2 \gtrless 0$ separately. Interestingly, the fit of all data gives $b_2 \simeq 0.9$, i.e., the function has a pole at $u^2 \gtrsim 1$.

\begin{figure}
    \centering
    \includegraphics[width = 0.60\linewidth]{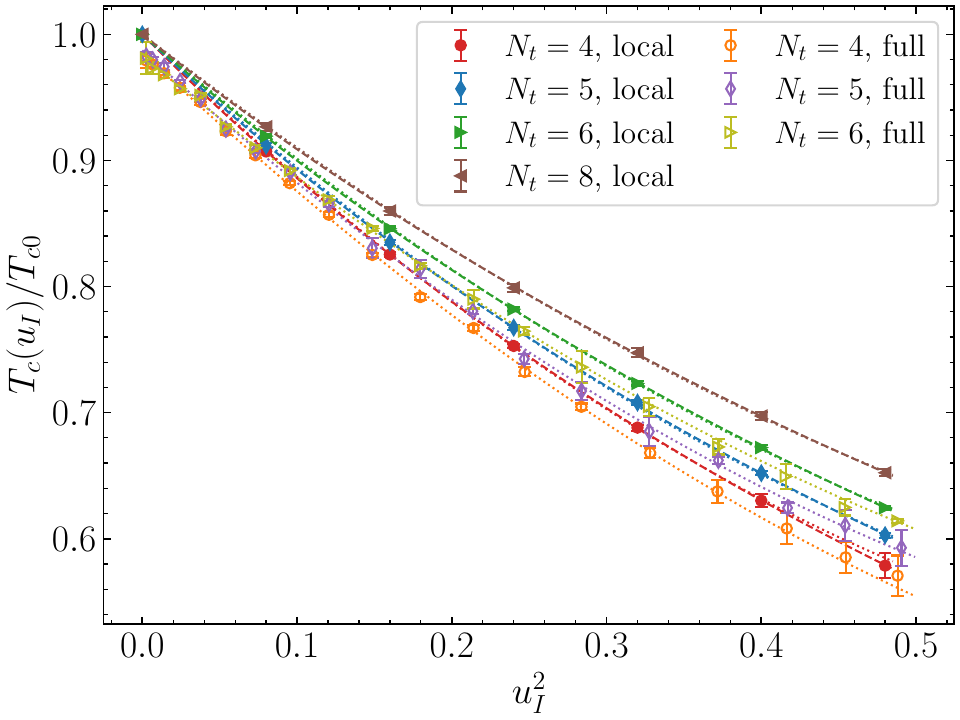}
    \caption{
    The (on-axis) critical temperature, calculated for the system with local action~\eqref{eq_S_E_continuum_local} (filled points), as a function of the $u_I$.
    The dotted (dashed) lines represent the best fit of the data by formula~\eqref{eq_fit_u_4} (by function~\eqref{eq_fit_u_4_rational}). The data are obtained on the lattices $4\times 16^3$, $5\times 20^3$, $6\times 24^3$, $8\times 32^3$. In addition, we show the critical temperatures calculated with the full action~(\ref{eq_S_E_continuum}) (empty points) on the lattices $4\times 24\times 145^2$, $5\times 30\times 181^2$, $6\times 36\times 217^2$ in Section \ref{sec_Coexistence}. }
    \label{fig_tcloc-u-compNt}
\end{figure}

In order to study how the local thermalization approximation works in the continuum limit, we calculated the critical temperature for the local action (\ref{eq_S_E_continuum_local}) on the lattices $4\times 16^3$, $5\times 20^3$, $6\times 24^3$ and $8\times 32^3$. The results are presented in Fig.~\ref{fig_tcloc-u-compNt}. In addition, we plot the critical temperatures calculated with the full action (\ref{eq_S_E_continuum}) on the lattices  $4\times 24\times 145^2$, $5\times 30\times 181^2$ and $6\times 36\times 217^2$ in Section \ref{sec_Coexistence}. For consistency, we consider the region $u_I^2 < 0.5$ for local action. It is seen from Fig.~\ref{fig_tcloc-u-compNt} that the results are in agreement with each other. Further, we fit the data in Fig.~\ref{fig_tcloc-u-compNt} by the function~\eqref{eq_fit_u_4} and~\eqref{eq_fit_u_4_rational} for each $N_t$ and extrapolate the coefficients to the continuum limit. Our results for the coefficients $k_2,~k_4$  are 
\begin{equation}
    k_2 = 0.869(31)\,,\qquad k_4 = 0.388(53) \,, 
    \label{eq_res_k24} 
\end{equation}
and the coefficients $c_2,~b_2$ take the values
\begin{equation}
    c_2 = 0.206(66)\,,\qquad b_2 = 0.694(101) \,. 
\label{eq_res_c2b2}
\end{equation}

Although the function with singularity better describes the data in regimes Im2  and Re2, we do not have any strong theoretical reasons to prefer one of these options for full action. Hence, the difference between the functions~\eqref{eq_fit_u_4} and~\eqref{eq_fit_u_4_rational} for $u^2 > 0$ may be considered as a systematic uncertainty of the analytic continuation procedure.

\subsection{Inconsistency with  the Tolman-Ehrenfest approach}

According to the Tolman-Ehrenfest (TE) law~\cite{Tolman:1930zza, Tolman:1930ona} in thermodynamic equilibrium in the rotating reference frame, the local temperature is given by the formula
\begin{equation}
    T(r) = \frac {T_0} {\sqrt{1 - (\Omega r)^2}} = \frac {T_0} {\sqrt{1 + (\Omega_I r)^2}} \,.
    \label{eq_TE}
\end{equation}
The first equality corresponds to the real {rotation}, while the second one corresponds to the imaginary rotation. 

The TE law (\ref{eq_TE}) suggests that real rotation effectively heats the system outside of the rotation axis, which naturally leads to the inhomogeneous phase in the rotating plasma in thermal equilibrium.
Indeed, the phase transition at some distance $r$ from the rotation axis is expected to occur when the effective local temperature $T(r)$ is equal to the critical temperature for the non-rotating system $T_{c0}$.
This assumption leads to the following prediction for local critical temperature:
\begin{equation}\label{eq_TE_local_tc}
    \frac{T_c(u)}{T_{c0}} = \sqrt{1 - u^2} \,.
\end{equation}
As a result, one can expect that for real rotation, deconfinement appears at the periphery and confinement at the center. This mixed-phase structure, described by Eq.~\eqref{eq_TE_local_tc}, was suggested in Ref.~\cite{Chernodub:2020qah} on the basis of a calculation in a low-dimensional confining model. Nevertheless, this theoretically motivated phase structure is in contradiction with the numerical simulations of this paper. 
Thus, a naive implementation of the TE law, which leads to formula~\eqref{eq_TE_local_tc}, does not describe correctly the properties of the rotating gluodynamics found in our first-principle numerical simulations. We conclude that the thermodynamics of rotating plasma is governed not only by the TE law but also by other factors that we discuss later.

In addition to the observed inconsistency with the TE prediction, there are many theoretical papers where it was shown that real rotation decreases the bulk-averaged critical temperature (see, for instance,~\cite{Chernodub:2016kxh, Chen:2020ath, Fujimoto:2021xix}). In view of these theoretical works, it is interesting why, within lattice simulation, the average over the volume critical temperature is increased by real rotation. 

To answer this question, take a look at the action (\ref{eq_S_E_continuum_local}). We have already found that the linear in the {imaginary frequency} $\Omega_I$ term does {not affect significantly} the critical temperature, and, {therefore}, it can be disregarded. The action (\ref{eq_S_E_continuum_local}) without the linear term can be written in the following form
\begin{align}
    S_{G} = \int\!  d^{4}x \ \Big[ & \beta \left( 
    (F^a_{x \tau})^2 + (F^a_{y \tau})^2 + (F^a_{z \tau})^2  + (F^a_{x z})^2 \right ) 
    + \tilde \beta \left ( (F^a_{y z})^2 + (F^a_{x y})^2 \right ) 
    \Big]\,,
    \label{eq_S_asym}
\end{align}
where the effect of rotation is encoded in the effective couplings
\begin{align}
    \tilde \beta = (1 - \Omega^2 r_0^2) \beta \equiv  (1 + \Omega_I^2 r_0^2) \beta\,, 
    \qquad 
    \beta = \frac{1}{2 g_{YM}^2}\,.
    \label{eq_betas}
\end{align}
The coupling $ \tilde \beta = \tilde \beta(r_0)$ depends on the distance $r_0$ from the rotation 
axis, thus encoding the effect of the curved metric in the co-rotating coordinates.

Thus, the external gravitational field~\eqref{eq_metric}, associated with the rotational motion, generates asymmetry in the coupling constants $\tilde \beta$ and $\beta$ of different components of the field-strength tensor squared $(F_{\mu\nu})^2$ that enter the action~\eqref{eq_S_asym}. The thermodynamic effect of the asymmetry in couplings~\eqref{eq_betas} can be assessed in the lattice gauge theory, which reveals that the anisotropy generates a rather peculiar effect: it leads to a change in the lattice spacing in the spatial and temporal directions~\cite{Karsch:1982ve}, and consequently, can affect the critical temperature of the deconfinement transition. While the discussed gluon action~\eqref{eq_S_asym} differs from the one considered in Ref.~\cite{Karsch:1982ve}, the remarkable effect of the action anisotropy can indeed be seen from our numerical results on the critical temperature shown in Fig.~\ref{fig_tcu-Im2Re2-check-continuation}. In an analytical form, this effect is excellently reproduced by the rational function~\eqref{eq_fit_u_4_rational} with the fitting parameters~\eqref{eq_res_c2b2}. 

Figure~\ref{fig_tcu-Im2Re2-check-continuation} indicates that the larger the asymmetry $\tilde \beta/\beta$,  the smaller the critical temperature. Taking into account that, for imaginary rotation, $\Omega_I$, the asymmetry of the couplings~\eqref{eq_betas} increases with the rise in the distance $r_0$ from the rotation action, ${\tilde \beta}/\beta = 1 + \Omega_I^2 r_0^2$, we arrive to the immediate conclusion that the periphery of the system transits to the deconfinement phase earlier than the central regions, and the average critical temperature decreases with imaginary rotation. These arguments remain true for real rotation, but in this case, they work in the opposite direction: for real rotation $\tilde \beta/\beta<1$, the confinement/deconfinement phases take place in the periphery/center regions, and the averaged critical temperature is increased. 

It should be noted here that the asymmetry generated by the external gravitational field is related to the magnetovortical term~\eqref{eq_S2_magn} in the gluon action. In the absence of this term, the mixed phase structure would be in a qualitative agreement with the TE prediction (see the plot for Im1-regime in Fig.~\ref{fig_imshowPLxy-v4-lambdas}).
Notice also that the same asymmetry terms, which are associated with the chromomagnetic fields, give rise to the negative moment of inertia measured in papers~\cite{Braguta:2023yjn, Braguta:2023kwl, Braguta:2023qex}. 

Thus, we conclude that the Tolman-Ehrenfest law is not the only factor that determines the thermodynamics of the rotating gluon plasma. The additional factor is the external gravity that generates the asymmetry in the gluon action, which, in turn, influences the dynamics of gluon matter. This mechanism cannot be accounted for by the simple Tolman-Ehrenfest formula (\ref{eq_TE}).
Therefore, the Tolman-Ehrenfest effect, if applied straightforwardly, would lead to a qualitatively wrong prediction for the behavior of the inhomogeneous phase of gluon plasma. In this sense, the Tolman-Ehrenfest picture is ``violated'' in the rotating gluon plasma.

\section{Discussion and conclusion}\label{sec_Conclusions}

Our work focuses on the lattice investigation of the confinement/deconfinement phase transition in rigidly rotating gluon plasma. To incorporate the rotation in the system, we employed a reference frame that is co-rotating with the gluon plasma, effectively describing the rotational effects by an external gravitational field with a suitable metric. Since direct lattice simulations of the full gluonic action with nonzero angular velocity are hindered by the sign problem, we performed numerical simulations using an imaginary angular velocity. The results at real angular velocities are subsequently recovered via an analytic continuation.

We have found that there appears to be a mixed-phase state for certain simulation parameters, thus confirming and refining our earlier studies~\cite{Braguta:2023iyx}. For \textit{imaginary} rotation in this state, the confinement phase is located in the center region, whereas the deconfinement phase is localized in the periphery of the studied volume. The mixed-phase state can be realized for on-axis temperatures smaller than the critical temperature of the non-rotating gluodynamics $T_{c0}$. For the higher temperatures $T>T_{c0}$, the mixed inhomogeneous phase does not appear. 

To describe the inhomogeneous phase, we introduced the local (pseudo)critical temperature $T_c(r)$. This is the temperature on the rotation axis, for which the system, with imaginary angular velocity $\Omega_I$, undergoes the confinement/deconfinement phase transition at the distance~$r$. From the construction, it is clear that for a given temperature $T = T_c(r)$, one has the confinement phase at distances smaller than $r$ and the deconfinement phase at larger distances.

We calculated the local critical temperature for various parameters of the simulations from the peak of the local Polyakov loop susceptibility. Disregarding the finite volume effects, which do not play an important role in the bulk of the studied volume, our results can be well described by the formula
\begin{align}\label{eq_conclusion1}
 \frac{T_c(r)} {T_{c0}} = 1 - \kappa_2 \left ( \Omega_I r \right )^2, 
 \phantom{\rm real} {\rm [imaginary\ rotation]},
\end{align}
where $r$ is the distance from the boundary to the rotation axis. The vortical curvature $\kappa_2$ weakly depends on the parameters of the simulations, in particular, on boundary conditions, and seems to be a universal quantity. The results imply that the local critical temperature on the axis of rotation $T_c(r=0)$ within the uncertainty coincides with the critical temperature of non-rotating gluodynamics $T_{c0}$. 
For the system with open boundary conditions, we also measured the sub-leading term, $\sim \kappa_4 r^4$, (see Eq.~\eqref{eq_fit_4_result} with the coefficients~\eqref{eq_res_kap24}), but we expect that the value of coefficient $\kappa_4$ strongly depends on the boundary conditions and it gives a minor contribution in bulk.

Formula (\ref{eq_conclusion1}) allowed us to carry out analytical continuation from imaginary to real rotation through the substitution $\Omega^2_I \to - \Omega^2$:
\begin{align}\label{eq_conclusion2}
 \frac {T_c(r)} {T_{c0}} = 1 + \kappa_2 \left ( \Omega r \right )^2,
 \phantom{\rm imaginary} {\rm [real\ rotation]}, 
\end{align}
which explicitly indicates that the local critical temperature growths increase both in the distance from the rotation axis $r$ and the angular velocity $\Omega$. The coefficient $\kappa_2$ in Eqs.~\eqref{eq_conclusion1} and \eqref{eq_conclusion2} is given in \eqref{eq_res_kap2}. Using Eq.~\eqref{eq_conclusion2}, we concluded that the mixed-phase state can be realized for \textit{real} rotation. However, in this case, the inhomogeneous phase appears at temperatures above the $T_{c0}$, and the deconfinement phase is localized in the central region while the confinement phase {appears to be} in the periphery of the system.
Confirming the earlier arguments put forward in Ref.~\cite{Braguta:2023yjn}, the radial dependence of the local critical temperature~\eqref{eq_conclusion2} does not demonstrate the behavior suggested by the Tolman-Ehrenfest law in the rotating gluon matter.

Although in our paper, we studied gluodynamics, one can expect that our qualitative results will not change if the dynamical fermions are accounted for in the simulation. This conclusion can be drawn from the results of paper~\cite{Braguta:2022str}  where it was shown that in rotating QCD, the gluon sector of the theory plays the major role, while the contribution of dynamical fermions is not significant.  

The action of rotating gluodynamics is a polynomial of the second power of angular velocity. In view of this observation, the gluon operators that contribute to the linear and quadratic in the $\Omega_I$ terms can be studied separately, i.e., one can switch on/off different operators and study the action in this regime. 
These operators have clear physical meaning. The linear operator is related to the total angular momentum of gluons. The quadratic one is related to the squares of the chromomagnetic fields. It is also interesting to note that the quadratic term is responsible for the negative values of the moment of inertia of gluon plasma~\cite{Braguta:2023yjn, Braguta:2023kwl, Braguta:2023tqz, Braguta:2023qex} below the supervortical temperature $T_s \simeq 1.5 T_{c0}$. 

Taking into account these interesting properties of the linear and quadratic operators, we have studied their influence on the mixed-phase state and local critical temperature. We found that the linear term generates the inhomogeneous phase transition in rotating gluodynamics, but its properties are opposite to that of the full theory. In particular, for \textit{imaginary} rotation, the linear operator gives rise to the deconfinement in the center and the confinement in the periphery region. The local critical temperature increases with $\Omega_I r$. The properties of the quadratic operator are similar to that of the full theory. We also found that the linear term plays a minor role in the dynamics of the full system, and the properties of the mixed-phase state are determined by the quadratic magnetovortical term. 

Lattice simulation of rotating gluodynamics without linear term is not spoiled by the sign problem. Therefore, one can conduct the lattice simulations for both real and imaginary angular velocity and study how analytic continuation works. Our results for the critical temperature and the distribution of the Polyakov loop suggest that within {a small numerical} uncertainty, the simulations with imaginary and real angular velocity are connected through the analytical continuation procedure~$\Omega^2 \leftrightarrow -\Omega_I^2$. 

In the last section of our work, we proposed and studied the local thermalization hypothesis. The idea is that for sufficiently slow rotation, the coefficients of the operators in the gluon action, which depend on the $\Omega_I r$, remain approximately the same if one varies the distance by a few correlation lengths of gluodynamics. Therefore, instead of the direct simulation of the whole inhomogeneous system, one can study homogeneous gluon action with the {anisotropic couplings} fixed at some distance. This approximation has valuable advantages over the direct simulation of rotating gluodynamics. In particular, there is no need to {perform} lattice simulations {at huge spatial} volumes. In addition, {in the homogeneous case}, there is no systematic uncertainty caused by the boundary effects. 

We have conducted a lattice study of the local thermalization hypothesis and found that the results are in good agreement with that of the direct simulation of rotating gluodynamics. Furthermore, in the local thermalization approximation, it is possible to observe the $\sim (\Omega_I r)^4$ contribution to the local critical temperature, which is not visible in the direct simulations because of the boundary effects. 
The results for local critical temperature in this approximation are well described by Eq.~\eqref{eq_fit_u_4} with the continuum limit coefficients~\eqref{eq_res_k24}.

Finally, the local thermalization approximation allows us to show that the external gravitational field of the rotating reference frame gives rise to asymmetry in the coupling constants of the gluon action. In turn, this leads to the shift of the critical temperature in the periphery regions as compared to the central {regions of the rotating system} and, as a result, to the appearance of the mixed inhomogeneous confinement-deconfinement phase.

\vskip 1mm
\paragraph*{\bf Summary of our results.}
For the convenience of the reader, we briefly summarize the results of this paper:
\begin{enumerate}
    \item The critical temperature of the deconfining transition on the rotating axis $T_c(r{=}0)$ for rotating gluonic matter coincides with the critical temperature of non-rotating gluonic matter $T_{c0}$. In other words, rotation produces no effect on the critical deconfining temperature on the rotating axis, thus affirming our earlier finding of Ref.~\cite{Braguta:2023iyx}.
    \item The local critical temperature $T_c(r)$ in the rotating system is mainly determined by the local velocity $\Omega r$ (see Section~\ref{sec_Coexistance-localTc}).
    This allowed us to introduce the approximation of local thermalization, which gave consistent results for $T_c(r)$ (see Section~\ref{sec_Local_results}).
    \item In agreement with Ref.~\cite{Braguta:2023iyx}, we confirm the existence of a novel inhomogeneous phase in the rotating Yang-Mills theory. Gluon matter rotating with a real-valued angular velocity resides in the deconfinement phase close to the axis of rotation surrounded by the confinement phase at larger radii. At a given temperature $T$ at the rotation axis, the deconfinement phase extends up to the radius $r$ given by the solution of the equation $T = T_c(r)$.
    \item The emergence of this exotic inhomogeneous phase appears to be a result of the quadratic magnetovortical coupling of the angular velocity to the magnetic component of the gluonic field. The standard linear coupling to the mechanical angular momentum plays a subleading role [{\it cf.} Eqs.~\eqref{eq_L_G_decomposition} and \eqref{eq_action_decomposition}].
    \item The gluon matter with the magnetovortical coupling is not affected by the sign problem, so it can be simulated with real-valued angular velocity. We confirm the crucial role of the magnetovortical coupling in the formation of the new inhomogeneous phase. We also demonstrate the validity of the analytical continuation of our results from imaginary to real-valued angular frequencies. 
    \item We show that the source of the disagreement of our first-principle result with a naive prediction suggested by the Tolman-Ehrenfest law is the result of the magnetovortical coupling of rotation to the magnetic gluonic condensate. This term is produced by the anisotropy of the gluonic action in curved spacetime in a co-rotating non-inertial reference frame of rotating plasma.
\end{enumerate}

\begin{acknowledgments}
The work of VVB, YaAG and AAR has been carried out using computing resources of the Federal collective usage center Complex for Simulation and Data Processing for Mega-science Facilities at NRC ``Kurchatov Institute'', \href{http://ckp.nrcki.ru/}{http://ckp.nrcki.ru/} and the Supercomputer ``Govorun'' of Joint Institute for Nuclear Research. The work of VVB and AAR, which consisted of the lattice calculation of the observables used in the paper and interpretation of the data,  was supported by the Russian Science Foundation (project no. 23-12-00072). The work of MNC was funded by the EU’s NextGenerationEU instrument through the National Recovery and Resilience Plan of Romania - Pillar III-C9-I8, managed by the Ministry of Research, Innovation and Digitization, within the project entitled ``Facets of Rotating Quark-Gluon Plasma'' (FORQ), contract no.~760079/23.05.2023 code CF 103/15.11.2022. The authors are grateful to Victor Ambru\cb{s}, Matteo Buzzegoli, Andrey Kotov and Dmitrii Sychev for useful discussions. 

\end{acknowledgments}

\appendix

\section{Width of the transition region}\label{app_width}

\begin{figure}[t]
    \centering
    \includegraphics[width = 0.49\linewidth]{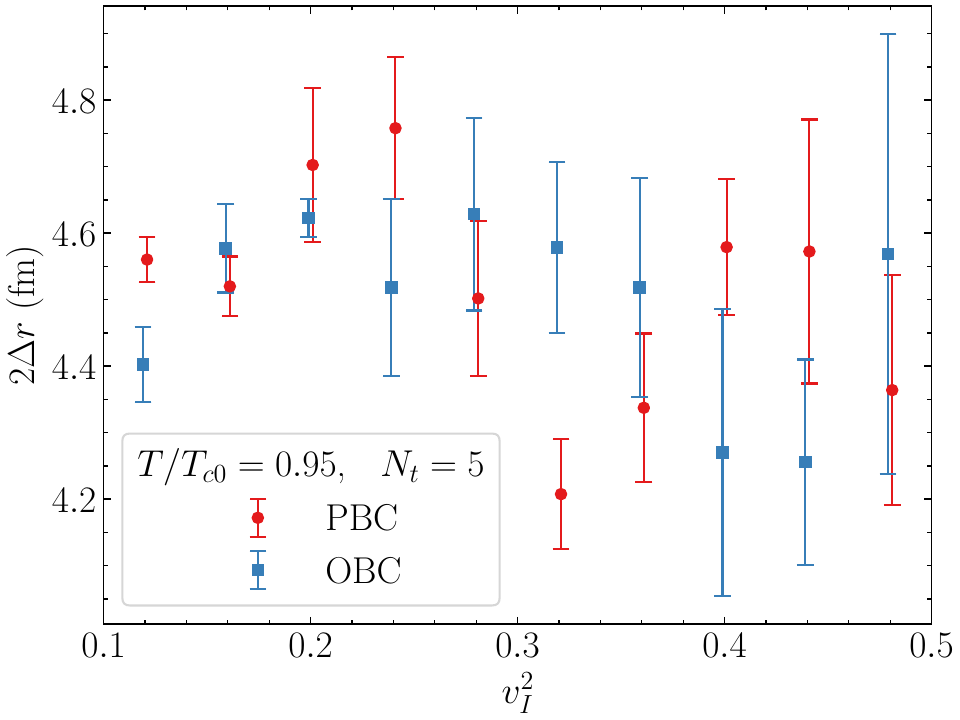}
    \hfill
    \includegraphics[width = 0.49\linewidth]{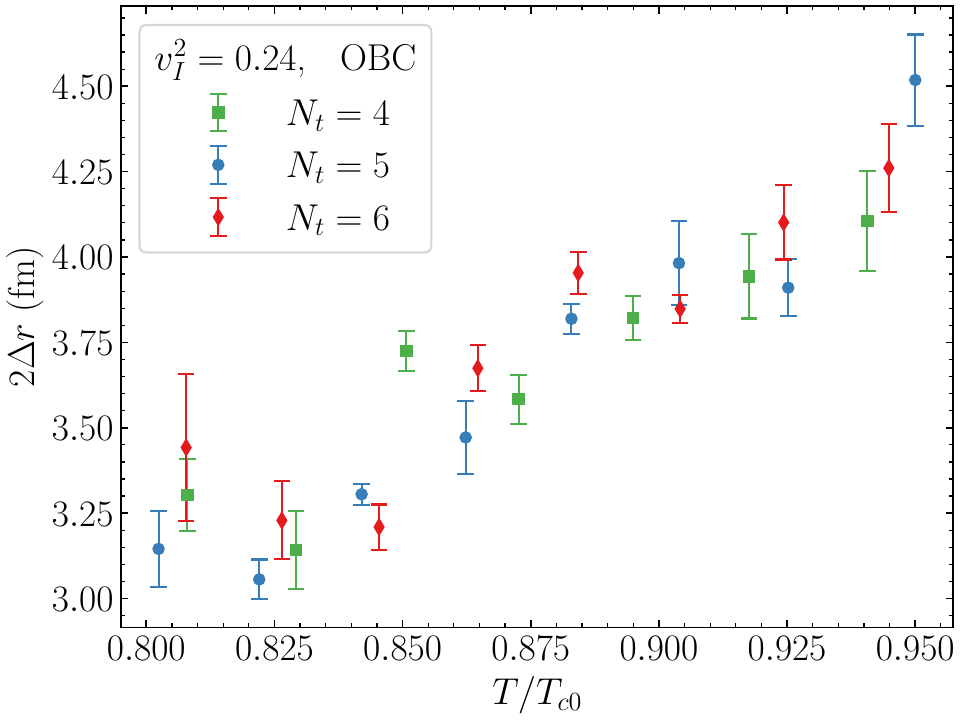}
    \caption{The spatial width $2\Delta r$ of the transition region as a function of the velocity of rotation for a fixed temperature $T/T_{c0} = 0.95$
    (left),
    and as a function of temperature for a fixed velocity of rotation $v_I^2 = 0.24$ 
    (right).
    The results are obtained on the lattices $4\times 24\times 145^2$, $5\times 30\times 181^2$ and $6\times 36\times 217^2$.
    }
    \label{fig_width}
\end{figure}

To estimate the spatial width of the transition region, we consider the radial dependence of the local Polyakov loop $L(r)$, defined by Eq.~\eqref{eq_polyakov_loop_xy}. In the vicinity of the transition, we fit the data by the following function:
\begin{equation}
    L(r) = A_0 + A_1 \tanh\left(\frac{r-r_0}{\Delta r} \right)\,,
\end{equation}
where $\Delta r$ is a half-width of the spatial transition. Note that here, we do not introduce any additional averaging over cylindrical layers and we do use values of the local Polyakov loop calculated in all spatial points in $x,y$-plane. The transition width $2\Delta r$ is approximately $\sim 3-5$~fm with a mild dependence on the rotation velocity and the temperature (see Fig.~\ref{fig_width}).

\begin{figure}[t]
    \centering
    \includegraphics[width = 0.6\linewidth]{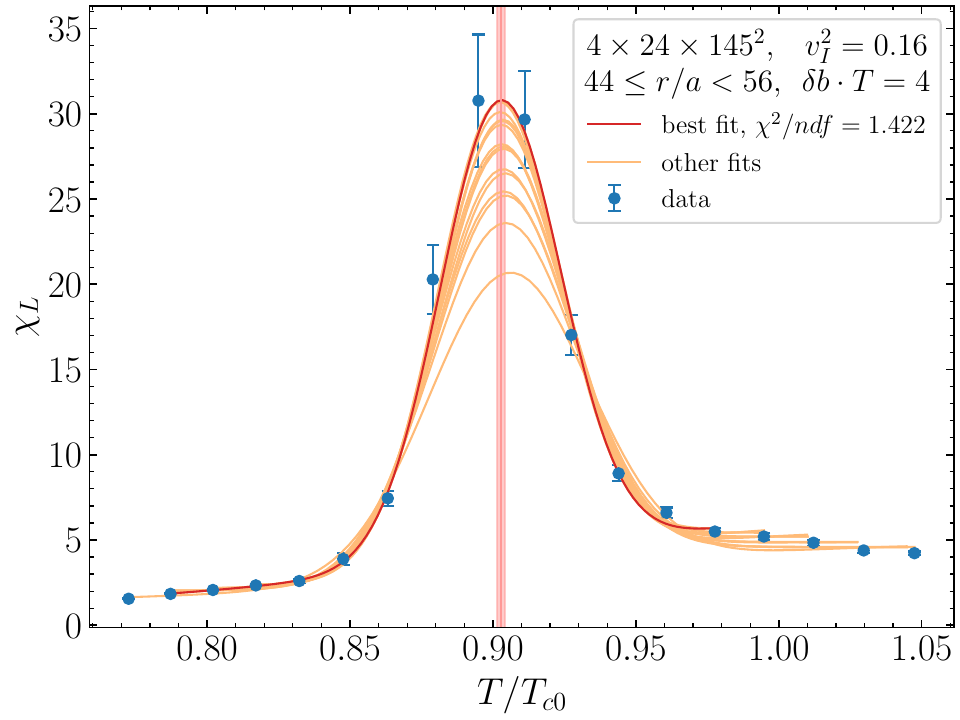}
    \caption{The temperature dependence of the susceptibility of the Polyakov loop for a given circular layer $44\leq r/a < 56$ of width $\delta r \cdot T = 3$ with boundary offset $\delta b \cdot T = 4$. The results of Gaussian fit~\eqref{eq_fit_gauss_tanh} for different subsets of data points are shown by red and orange lines. The best fit with $\chi^2/{\rm n.d.f.} = 1.422$ is shown by the red line. Orange lines represent the fit results, which are used to estimate systematic uncertainty. The vertical shaded region denotes the uncertainty of the local critical temperature for this circular layer, which includes both statistical and systematic contributions.
    The data are obtained on the lattice $4\times 24\times 145^2$ with OBC.}
    \label{fig_fit_chi_systematics}
\end{figure}

\section{Calculation of the local critical temperature }\label{app_Fit}

For a given circular layer of the width $\delta r$ with an average radius $r$, we calculate a local Polyakov loop susceptibility~\eqref{eq_usceptibility} for different values of the (on-axis) temperature. The resulting data are fitted by the Gaussian function of a general form
\begin{equation}\label{eq_fit_gauss_tanh}
    \chi_L(T) = c_0 + c_1 \exp\left(-\frac{(T - T_{c})^2}{2\, c_2^2} \right) + 
    c_3 \tanh \left( \frac{T-T_{c}}{c_4} \right)\,,
\end{equation}
what gives us an estimation of the corresponding local critical temperature $T_c$. {The role of the last term in Eq.~\eqref{eq_fit_gauss_tanh} is to make the fitting function compatible with} different {asymptotic} values below and above the critical temperature $T_c$.

To estimate the systematic uncertainty, we fit the data using different subsets of data points (at least 12 data points are always used for a 6-parameter fit). As the final result for the local critical temperature, we use the fit coefficient $T_c$ from the subset, which provides the best-fit quality ($\chi^2/{\rm n.d.f.}$ is usually about $\sim 1-3$).
An example of this procedure is illustrated in Fig.~\ref{fig_fit_chi_systematics}.

\section{Analysis of the finite-volume and lattice spacing effects}\label{app_NsNt}

The results for local critical temperature $T_c(r)$, which were {presented} in Section~\ref{sec_Coexistance-localTc}, are well described in the whole range of available data by the quartic function
\begin{equation}\label{eq_fit_4_app} 
    \frac{T_c(r)}{T_{c0}} = C_0 - C_2 \left( \frac{r}{R}\right)^2 + C_4 \left( \frac{r}{R}\right)^4\,.
\end{equation}
Note that the first two terms in Eq.~\eqref{eq_fit_4_app} are not sufficient to describe all data points.
However, the quadratic fitting function (i.e., Eq.~\eqref{eq_fit_4_app} with $C_4=0$) can be used if we cut the fit range at $(r/R)^{\rm (max)} \sim 0.5$. In this case, fit quality becomes appropriate $\chi^2/{\rm n.d.f.}\sim 1-3$. The quadratic fit was {also} used in Ref.~\cite{Braguta:2023iyx}.

\begin{figure} 
    \centering
    \includegraphics[width = 0.60\linewidth]{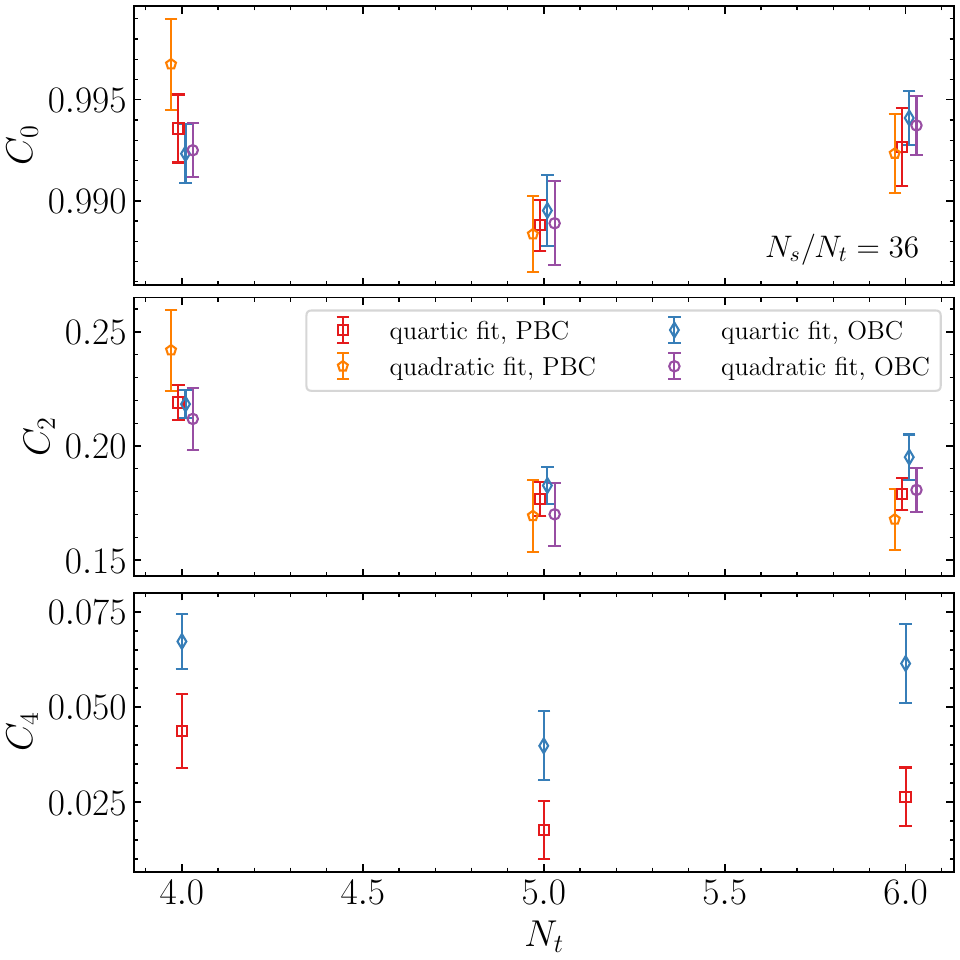}
    \caption{
    The coefficients $C_0$, $C_2$, $C_4$ as a function of the $N_t$.
    The results are obtained on the lattices $4\times 24\times 145^2$, $5\times 30\times 181^2$ and $6\times 36\times 217^2$ at the fixed rotating velocity $v_I^2 = 0.16$ for open and periodic boundary conditions.}
    \label{fig_coef-compBC-compNt}
\end{figure}

To study the effects of finite lattice spacing, we check the results for the lattices with $N_t = 4,\,5,\,6$ and the same aspect ratios for fixed boundary velocity $v_I^2 = 0.16$.
The comparison of the fit coefficients for the lattices $4\times 24\times 145^2$, $5\times 30\times 181^2$, $6\times 36\times 217^2$ with open and periodic boundary conditions are presented in Fig.~\ref{fig_coef-compBC-compNt}. The uncertainties include both statistical and systematic contributions, as it is described in Appendix~\ref{app_Systematic}.

From Fig.~\ref{fig_coef-compBC-compNt}, one can conclude that the results for two higher coefficients $C_0,\, C_2$ are almost independent of boundary conditions, while the coefficient $C_4$ are slightly differ for PBC and OBC. Note that the results for the coefficients $C_0$ and $C_2$ from the quartic and quadratic fit are in reasonable agreement with each other. The results are mildly dependent on the lattice spacing.

\begin{figure}[t]
    \centering
    \includegraphics[width = 0.60\linewidth]{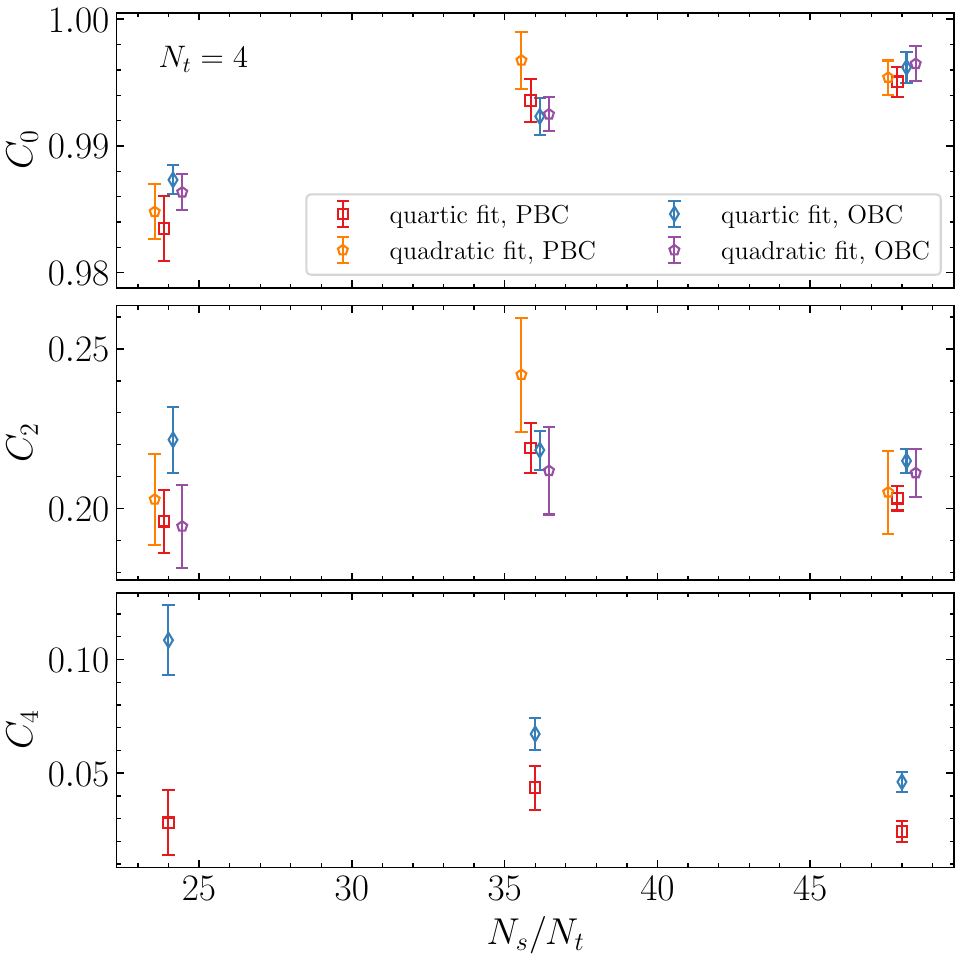}
    \caption{
    The coefficients $C_0$, $C_2$, $C_4$ as a function of transversal lattice size $N_s/N_t$. The results are obtained on the lattices $4\times 24\times 97^2$, $4\times 24\times 145^2$, $4\times 24\times 193^2$ with open and periodic boundary conditions at the fixed rotating velocity $v_I^2 = 0.16$.}
    \label{fig_coef-compBC-compNs}
\end{figure}

The results for the local critical temperature, presented above, were calculated on the lattices with the transversal aspect ratio $N_s/N_t = 36$. To study the dependence of the result on the {spatial size of the system}, we calculate the local critical temperatures using the lattices with $N_s/N_t = 24,\,48$ and the same value of boundary velocity $v_I^2 = 0.16$. A comparison of the fit coefficients for the lattices $4\times 24\times 97^2$, $4\times 24\times 145^2$, and $4\times 24\times 193^2$ are presented in Fig.~\ref{fig_coef-compBC-compNs}.

From Fig.~\ref{fig_coef-compBC-compNs}, it follows that the coefficient $C_0$, which {determines} the local critical temperature at the axis of rotation, tends to unity with increasing of the system size $R \propto N_s$. It means that a small deviation of $T_c(0)$ from $T_{c0}$ may be associated with the effects of finite system radius. The second coefficient, $C_2$, is almost independent of $N_s$ and boundary conditions within the uncertainty, while the next coefficient, $C_4$, is different for open and periodic boundary conditions. This coefficient, $C_4$, is larger in the case of open boundary conditions, but it decreases with the growth of the lattice size. {As the size of the lattice increases, the results for OBC and PBC approach each other.} This result means that the effects of the boundary conditions also become smaller as the system radius increases.

\section{Estimation of systematic uncertainty for the fit coefficients}\label{app_Systematic}

\begin{figure}[t]
    \centering
    \includegraphics[width = 0.60\linewidth]{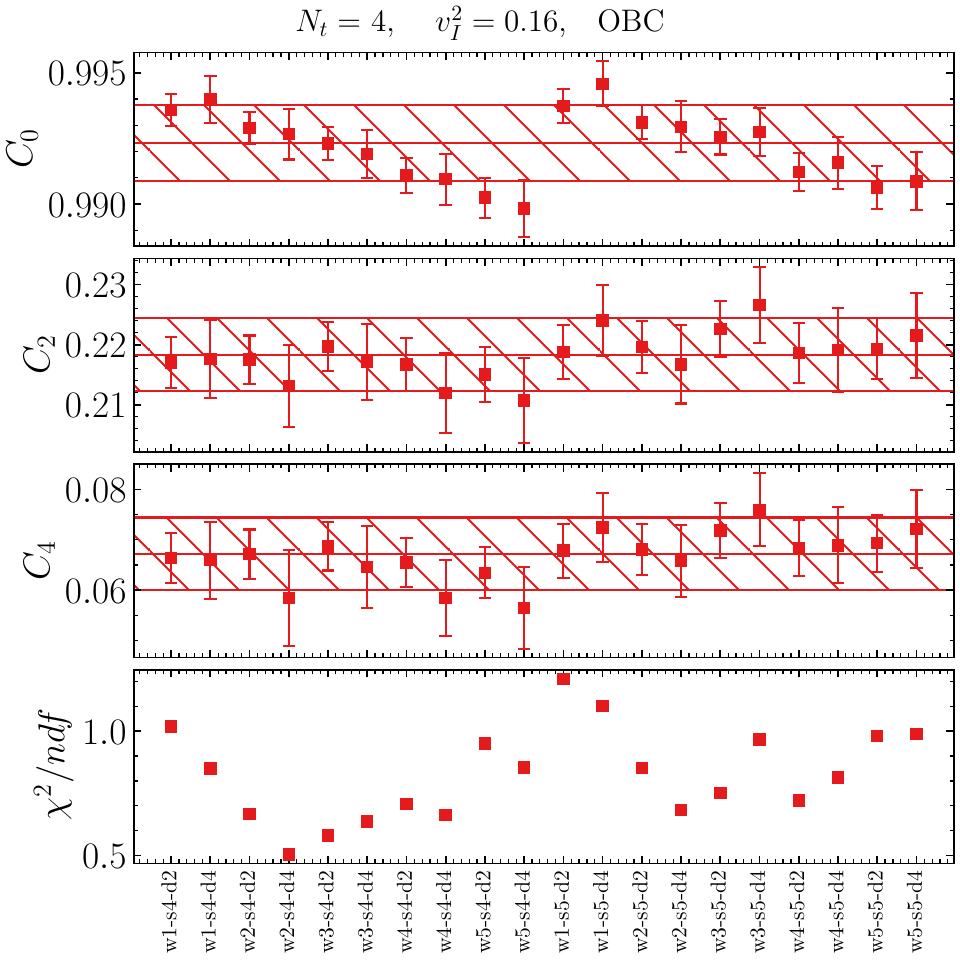}
    \caption{The coefficients $C_0$, $C_2$, $C_4$ of the best fit of the local critical temperature by the function~\eqref{eq_fit_4}, and the fit quality $\chi^2/\text{{\rm n.d.f.}}$, obtained with different widths of circular layer $\delta r\cdot T$ (w), boundary skips $\delta b\cdot T$ (s), and distance between the inner radius of used neighboring layers (d). The horizontal lines with hatched uncertainty show the final results. The data are obtained on the lattice $4\times 24\times 145^2$ with OBC at $v_I^2 = 0.16$.
    }
    \label{fig_Ci_systematics}
\end{figure}

We check how the fit coefficients $C_0$, $C_2$, and $C_4$ depend on the averaging width, and no significant dependence has been found. A minor dependence for the coefficient $C_0$ is observed (this dependence is more or less profound depending on the lattice size and the rotation velocity). This coefficient becomes closer to unity with the decrease of the layer width $\delta r$. At the same time, other coefficients {are not sensitive to the averaging width.} So, we do not take $\delta r \to 0$ limit of the coefficients and consider the choice of the circular layer width $\delta r$ as a source of additional systematic error. We average the fit coefficients $C_i$ overall considered values $\delta r \cdot T = 1,\,2,\,3,\,4,\,5 $ and use this value as the result for a given lattice and rotational velocity $v_I$.

Note that the neighboring circular layers may partially intersect {with each other}. To check the stability of the results with respect to the distance between neighboring layers, we fit the data from layers shifted by different steps of 2 and $N_t$ lattice sites along the radius and add these results to the averaging.

Another source of the statistical ambiguity is associated with the value of the boundary skip $\delta b$ (for the quartic fit) or fit range $x = r/R < x_{\rm max}$ (for the quadratic fit). We check how the fit coefficients vary with minor changes in this parameter ($x_{\rm max} = 0.4,\, 0.5$ or $\delta b = 4,\,5$) and also include this variation into the systematic uncertainty.

So, the final results for the fit coefficient include all systematic uncertainties discussed above, as well as a statistical one. An example of this calculation is shown in Fig.~\ref{fig_Ci_systematics} for the lattice $4\times 24\times 145^2$ with OBC at $v_I^2 = 0.16$.

\bibliographystyle{JHEP}
\bibliography{plasma.bib}



\end{document}